\documentclass[%
twocolumn,
 aip,
 amsmath,amssymb,
 reprint,%
]{revtex4-1}
\usepackage{float}
\usepackage{graphicx}% Include figure files
\usepackage{dcolumn}% Align table columns on decimal point
\usepackage{bm}% bold math
\usepackage[utf8]{inputenc}
\usepackage[T1]{fontenc}
\usepackage{mathptmx}
\usepackage{etoolbox}
\usepackage{mathtools}
\usepackage{amsmath}
\usepackage{caption}
\usepackage{subcaption}
\usepackage[T1]{fontenc}
\usepackage[utf8]{inputenc}
\usepackage{xcolor}

\makeatletter
\def\@email#1#2{%
 \endgroup
 \patchcmd{\titleblock@produce}
  {\frontmatter@RRAPformat}
  {\frontmatter@RRAPformat{\produce@RRAP{*#1\href{mailto:#2}{#2}}}\frontmatter@RRAPformat}
  {}{}
}%
\makeatother
\begin{document}

\preprint{AIP/123-QED}

\title[]{Limits of Entrainment \\ of Circadian Neuronal Networks}
% Force line breaks with \\
\author{Yorgos M. Psarellis}
\affiliation{Department of Chemical and Biomolecular Engineering, Johns Hopkins University.}%Lines break automatically or can be forced with \\
\author{Michail Kavousanakis}%
\affiliation{School of Chemical Engineering, National Technical University of Athens.
%Authors' institution and/or address%\\This line break forced with \textbackslash\textbackslash
}%

\author{Michael A. Henson}
\affiliation{%
Department of Chemical Engineering, University of Massachusetts, Amherst%\\This line break forced% with \\
}%

\author{Ioannis G. Kevrekidis *}
 \email{yannisk@jhu.edu.}
 \affiliation{%
 Department of Chemical and Biomolecular Engineering, Johns Hopkins University.} 
 \affiliation{Department of Applied Mathematics and Statistics, Johns Hopkins University.}

\date{\today}% It is always \today, today,

\begin{abstract}
Circadian rhythmicity lies at the center of various important physiological and behavioral processes in mammals, such as sleep, metabolism, homeostasis, mood changes and more. It has been shown that this rhythm arises from self-sustained biomolecular oscillations of a neuronal network located in the Suprachiasmatic Nucleus (SCN). Under normal circumstances, this network remains synchronized to the day-night cycle due to signaling from the retina. Misalignment of these neuronal oscillations with the external light signal can disrupt numerous physiological functions and take a long-lasting toll on health and well-being. In this work, we study a modern computational neuroscience model to determine the limits of circadian synchronization to external light signals of different frequency and duty cycle. We employ a matrix-free approach to locate periodic steady states of the high-dimensional model for various driving conditions. Our algorithmic pipeline enables numerical continuation and construction of bifurcation diagrams w.r.t. forcing parameters. We computationally explore the effect of heterogeneity in the circadian neuronal network, as well as the effect of corrective therapeutic interventions, such as that of the drug molecule Longdaysin. Lastly, we employ unsupervised learning to construct a data-driven embedding space for representing neuronal heterogeneity. 
\end{abstract}

\maketitle

\begin{quotation}
It is long known that the human brain not only keeps time, but adapts its own clock to the day-night (circadian) cycle. This is achieved when an internal, ``default'' rhythm synchronizes with the external, forcing cycle of alternating light and darkness. In this work, this synchronization phenomenon is computationally studied, using a detailed model of the group of neurons that constitute the circadian clock. We employ tools from nonlinear dynamics to estimate the \textit{limiting} conditions under which such synchronization is possible; and what happens when it is not. We also investigate how a certain drug molecule can affect the ability of the circadian clock to synchronize. The algorithms employed are well suited for the complexity of this problem: matrix free simulator-based solvers that efficiently handle high-dimensional equation systems and unsupervised learning techniques that  explore heterogeneity in such networks.
%find structure in large networks.

% The ``lead paragraph'' is encapsulated with the \LaTeX\ 
% \verb+quotation+ environment and is formatted as a single paragraph before the first section heading. 
% (The \verb+quotation+ environment reverts to its usual meaning after the first sectioning command.) 
% Note that numbered references are allowed in the lead paragraph.
% %
% The lead paragraph will only be found in an article being prepared for the journal \textit{Chaos}.
\end{quotation}

%%%%%%%%%%%%%%%%%%%%%%%%%%%%%%%%%%%%%%%%%%%%%%%%%%%%%%%%%%%%%%%

\section{\label{Sec:intro}Introduction}

The principal circadian pacemaker is located in the two Suprachiasmatic Nuclei of the anterior hypothalamus. The SCN directly receives input from photosensitive cells of the retina via the retinohypothalamic tract \cite{Ma2022-qr, Hastings2018} while it is also connected to the pineal gland, inducing melatonin production during the night \cite{Sack2007, Ma2022-qr}. The SCN also coordinates secondary cellular pacemakers across the body by controlling biochemical (e.g. neuroendocrine) signals that entrain them \cite{Hastings2018}.

This orchestrating role puts SCN neuronal networks at the center of numerous physiological processes and, consequently, its dysfunction gives rise to numerous disorders. Circadian desynchrony is linked to sleep disorders (jet lag disorder, shift work disorder,  advanced sleep phase disorder, delayed sleep phase disorder, free-running disorder, irregular sleep-wake rhythm) \cite{Sack2007,   10.1093/sleep1/30.11.1484},  cardiovascular disease, obesity, hypertension \cite{Scheer2009},
nephropathy \cite{kidney10.2337/diacare.14.8.707}, cancer \cite{Savvidis2012}, depression and bipolar disorder \cite{Germain2008, Vadnie2017}.

Many pharmacological interventions have been suggested to correct misalignments of the circadian rhythm with the external day-night signal. Benzodiazepines and melatonin have been popular choices \cite{Turek1986, melatonindoi:10.1056/NEJM200010123431503}. As the understanding of the the exact biochemical processes involved  deepens, more targeted therapies are developed. Small molecules like Longdaysin (CAS No. : 1353867-91-0) have been extensively studied ({\em in vitro} and {\em in silico}) as therapeutics in circadian rhythm disorders, as they directly intervene to the gene regulatory network giving rise to the oscillations \cite{Drugs_summary, StJohn2040, hirota2012}.

In this work, we use advanced algorithms from numerical analysis/scientific computation and  nonlinear dynamics to investigate the ability of SCN neurons and SCN neuronal networks to synchronize with the external day-night signal (in dynamics terms, with the external \textit{forcing}). Specifically, we explore entrained periodic solutions of {\em high dimensional} dynamical systems arising from computational biology models, with respect to different: 
\begin{enumerate}
    \item Forcing angular frequencies $\omega_f,$ where $\omega_f =  \frac{2\pi}{T_f},$ for a forcing period $T_f$ of the day-night (light-dark) cycle.
    \item Forcing duty cycles $\phi, $ where $\phi = \frac{T_{light}}{T_f}$ for a day duration of $T_{light}$.
    \item Simulated Longdaysin effects.
    \item Network Heterogeneity extents.
\end{enumerate}
For this purpose, several informative bifurcation diagrams are constructed and the extracted bifurcation points demarcate  the limits of entrainment of circadian neurons or neuronal networks. Moreover, we construct a reduced, data-driven ``emergent space'' description of neuronal behavior heterogeneity using unsupervised learning. 

The rest of the manuscript is organized as follows: Section \ref{Sec:model} includes all necessary information for the computational model used. Then, in Section \ref{Sec:algos} the main algorithms used are outlined and put into the context of circadian entrainment. Section \ref{Sec:results} presents our results, and Section \ref{Sec:conc} our conclusions. 

% \subsection{\label{sec:level2}Second-level heading: Formatting}

% This file may be formatted in both the \texttt{preprint} (the default) and
% \texttt{reprint} styles; the latter format may be used to 
% mimic final journal output. Either format may be used for submission
% purposes; however, for peer review and production, AIP will format the
% article using the \texttt{preprint} class option. Hence, it is
% essential that authors check that their manuscripts format acceptably
% under \texttt{preprint}. Manuscripts submitted to AIP that do not
% format correctly under the \texttt{preprint} option may be delayed in
% both the editorial and production processes.

% The \texttt{widetext} environment will make the text the width of the
% full page, as on page~\pageref{eq:wideeq}. (Note the use the
% \verb+\pageref{#1}+ to get the page number right automatically.) The
% width-changing commands only take effect in \texttt{twocolumn}
% formatting. It has no effect if \texttt{preprint} formatting is chosen
% instead.

\section{Computational Model}
\label{Sec:model}

\begin{figure*} %15:5
    \centering
    \makebox[0pt]{\includegraphics[width=15cm, height=5cm]{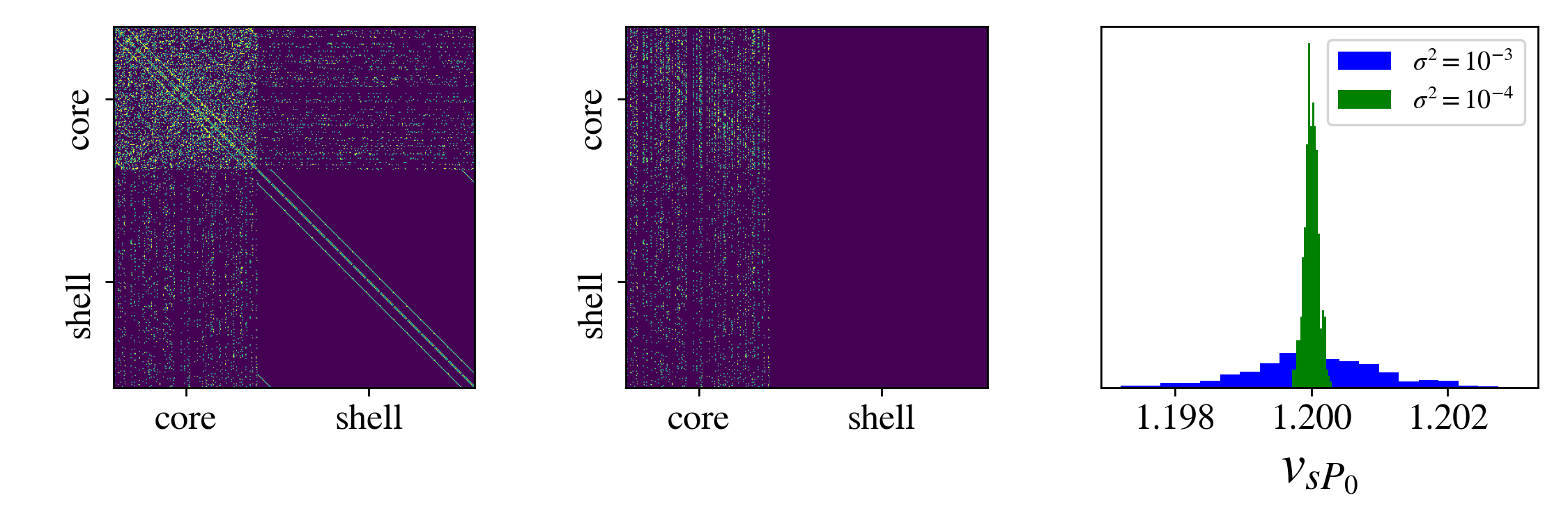}}    
    \caption{(left) GABA adjacency matrix, (middle) VIP adjacency matrix, (right) histograms sampled from different  $v_{sP_0}$ considered here.}
    \label{fig:inputs}
\end{figure*}

The computational model used here is adapted from the work of Vasalou et al. \cite{VASALOU201112, VASALOU201144}. They constructed a state-of-the-art computational model of circadian neuronal networks by coupling three components: 
\begin{enumerate}
    \item \textbf{Biomolecular clock}: The computational model of the biomolecular clock was developed by Leloup and Goldbeter   \cite{leloupgold2003, LELOUP2004541} and describes the regulatory loops involving the \textit{Per, Cry, Bmal1, Clock} and \textit{Erv-Erb$\alpha$}  genes.  The interaction of these loops gives rise to circadian oscillations.
    \item \textbf{Electrophysiology dynamics:} This component describes the membrane dynamics of circadian neurons and the way they are coupled with neurotransmitter signaling i.e. with $\gamma$-aminobutyric acid (GABA) and vasoactive intestinal polypeptide (VIP) signaling \cite{To2007, VASALOU201144}. The firing frequency is incorporated in the model by associating membrane voltage and ion conductances with circadian gene expression \cite{vasalou_1neuron}.
    \item \textbf{Network connectivity:}  An ensemble of 425 neurons was chosen for a realistic representation of the SCN circadian network \cite{VASALOU201144}. All neurons are linked via VIP and GABA neurotransmitter networks generated from a small-world architecture resembling neuron connectivities in the SCN \cite{VASALOU201144} (see Fig.\ref{fig:inputs} for example).
\end{enumerate}

Henson \cite{multi_henson_2013} provides a review comparing this model with others in the literature. There is also experimental evidence supporting the model's validity \cite{VASALOU201144,leloupgold2003}.

In this model, the day-night (light-dark) cycle is modeled as a step function with  period $T_f = T_{light} +  T_{dark}$ and duty cycle $\phi = \frac{T_{light}}{T_f}$. Light forcing is incorporated in the dynamical system implicitly, by assuming that during the light phase, the AMPA/NDMA and VPAC2 receptors are saturated (in term of model functions: $b_{Glu_R}(t)=1, b_{VIP}(t)=1$  for $t \in [0, T_{light}]$)    \cite{VASALOU201144}. In contrast to  Vasalou et al. \cite{VASALOU201144} we assume that the photic effect is uniform across all SCN (both core and shell) neurons.

The effect of Longdaysin can also be incorporated in the model.  Longdaysin is a small drug molecule acting as a casein kinase I (CKI) inhibitor. It is hypothesized to increase the time required before the \textit{Per-Cry} complex can enter the nucleus to repress transcription \cite{hirota2012, StJohn2040}. In our model, this effect can be simulated by different values of the parameter $k_1$, the nuclear entry rate of phosphorylated \textit{Per-Cry} complex \cite{StJohn2040}. Therefore, in our studies, increased values of Longdaysin translate to lower values of $k_1$. Note that, here, we assume that the concentration of Longdaysin is constant and not subject to pharmacodynamics.

The realization of the circadian network studied here includes heterogeneity in the  parameter $v_{sP_0}$, the basal transcription rate of the \textit{Per} mRNA, across the neurons. This  parameter has been shown to strongly affect the ability of circadian neurons to sustain intrinsic oscillations, while the rhythmic phenotype of \textit{Per} has been shown experimentally to vary \cite{yamaguchi2003}. Specifically, $v_{sP_0}$ has been sampled from $\mathcal{N}(1.2, \sigma^2)$, similarly to
\cite{VASALOU201112} (see Fig.\ref{fig:inputs} for example). It is important to mention, that for reproducibility, the random seed used to generate $v_{sP_0}$ and the small-world networks is always fixed.

% \begin{center}
%      \makebox[\textwidth][c]{\includegraphics[width=15cm, height=5cm]{Figures/inputs.png}}
%      \captionof{figure}{(left) GABA adjacency matrix, (middle) VIP adjacency matrix, (right) histograms sampled form different  $v_{sP_0}$ considered here.}
% \end{center}

Putting all components together, each circadian neuron is described by 21 Ordinary Differential Equations (ODEs), each of which describes the time evolution of a relevant chemical species. When the entire network is simulated, the resulting system is 8925-dimensional.

\section{Algorithms for computing synchronization of large oscillating networks}
\label{Sec:algos}

\subsection{ Stroboscopic Map }
The studied model is a periodically forced dynamical system, and computations are performed using a stroboscopic map, i.e., we evaluate the periodic solution at a selected, fixed phase of the forcing term \cite{KEVREKIDIS19861263}.
This acts as a natural pinning condition: the period of forced oscillations is, by necessity, an integer multiple of the period of the forcing term.
This is why the term
``entrainment'' (of the oscillator by the forcing) is used.

The dynamics of the problem are described by a system of ODEs, which has the general form:

\begin{equation} \label{eq:generalODE}
    \frac{\textrm{d} \bf{X}}{\textrm{d} t} = \mathbf{f} \left( \mathbf{X}(t);\lambda \right),
\end{equation}
    
\noindent where $\bf{X}$ denotes the state vector, and $\lambda$ is a parameter of the system. 
To compute a periodic solution at a given period $T_f$, we seek $\mathbf{X}(0)$ satisfying:
\begin{equation}\label{eq:periodicity}
    \mathbf{R} \coloneqq \mathbf{X}(0) - \mathbf{X}(T_f) = \mathbf{X}(0) - \boldsymbol{\Phi}_{T_f}\left(\mathbf{X}(0) ; \lambda \right),
\end{equation}
\noindent with $\boldsymbol \Phi_{T_f}$ denoting the temporal evolution operator of the system that reports the solution after a time interval $T_f$.

One converges computationally to a periodic solution employing the Newton-Raphson iterative technique, which solves at each iteration the linearized system:
    
\begin{equation} \label{eq:NR}
    \left [\mathbf{I} - \frac{\partial \boldsymbol{\Phi}_{T_f}}{\partial \mathbf{X}(0)} \right] \delta \mathbf{X}(0) = -\left[           \mathbf{X}(0) - \boldsymbol{\Phi}_{T_f} \left(\mathbf{X}(0) \right) \right] 
\end{equation}
    
\noindent and updates $\mathbf{X}(0)$ with $\mathbf{X}(0)+\delta \mathbf{X}(0)$ until convergence. 
Here, the computation of the variational matrix, $\mathbf{V}(t) \coloneqq \Large{\frac{\partial \boldsymbol{\Phi}_{T_f}}{\partial \mathbf{X}(0)}}$ requires the solution of the following set of ODEs:
    
\begin{equation} \label{eq:variational}
    \frac{\textrm{d} \mathbf{V}}{\textrm{d} t} = \mathbf{J} \cdot \mathbf{V},
\end{equation}
\noindent where $\mathbf{J}$ is the Jacobian of $\mathbf{f}$, $\mathbf{J} \coloneqq \frac{\partial \bf{f}}{\partial \bf{X}}$, and initial condition: $\mathbf{V}(t=0)=\mathbf{I}$.
When the Jacobian's analytical derivation is nontrivial, one can resort to numerical approximations by calling the time stepper $\boldsymbol{\Phi}_{T_f}$ at appropriately perturbed values of the unknowns.
This approach would be, however, impractical for large-scale problems.
For this reason, Eq.~\ref{eq:NR} is solved with matrix-free iterative solvers, such as the Newton-GMRES (Generalized Minimum Residual) method \cite{saad:gmres, Kelley2004, Dickson2007, Vandekerckhove2009}.

\subsection{Matrix-Free Newton-GMRES}
The decisive advantage of Newton-GMRES is that the explicit calculation and storage of the Jacobian of the fixed point problem is not required; instead, the action of the Jacobian on systematically selected vectors is approximated numerically. 
We only require matrix-vector multiplications, that can be estimated at low cost by calling $\boldsymbol{\Phi}_{T_f}$ from appropriately perturbed initial conditions \cite{Nkgmres_kelley, Siettos2003}.
In particular, to solve  Eq.~(\ref{eq:NR}) iteratively with GMRES, one estimates the directional derivative of $\boldsymbol{\Phi}_{T_f}$ on known vectors, $\mathbf{q}$:
     
\begin{equation} \label{eq:matvec}
    \frac{\partial \boldsymbol{\Phi}_{T_f}}{\partial \mathbf{X}(0)} \cdot \mathbf{q} \approx \frac{\boldsymbol{\Phi}_{T_f} \left( \mathbf{X}(0) + \epsilon \mathbf{q} \right)-\boldsymbol{\Phi}_{T_f} \left( \mathbf{X}(0) \right)}{\epsilon},
\end{equation}
\noindent where $\epsilon$ is a small and appropriately chosen scalar.
Such matrix-vector multiplications are required to construct the $j-$th Krylov subspace, which for this problem is:
     
\begin{equation}
    K_j = K_j ( \mathbf{A}, r_o ) = \textrm{span} \left\{ r_o, \mathbf{A} r_o,...,\mathbf{A}^{j-1}r_o \right \},
\end{equation}
     
\noindent where:
     
% \begin{equation}
%     \mathbf{A} = 
%     \mathbf{I} - \frac{\partial \boldsymbol{\Phi}_{T_f}}{\partial \mathbf{X}(0)}
% \end{equation}

% \noindent and
    
% \begin{equation}
%     r_o = - \left[
%     \mathbf{X}(0) - \boldsymbol{\Phi}_{T_f} \left(\mathbf{X}(0) \right) \right]
% \end{equation}
\begin{equation}
    \mathbf{A} = 
    \mathbf{I} - \frac{\partial \boldsymbol{\Phi}_{T_f}}{\partial \mathbf{X}(0)} , \hspace{0.2cm} r_o = - \left[
    \mathbf{X}(0) - \boldsymbol{\Phi}_{T_f} \left(\mathbf{X}(0) \right) \right],
\end{equation}

\noindent with $r_o$ being the initial error given that the initial guess to solve Eq.~(\ref{eq:NR}) is $\delta \mathbf{X}(0)^{(0)}=\mathbf{0}$.
In practice, GMRES applies the Arnoldi iteration \cite{Arnoldi1951} to find an orthonormal basis for $K_j$, by systematically evaluating matrix-vector products as described in Eq.~(\ref{eq:matvec}).
  
\subsection{Pseudo-arclength continuation}
    
To trace entire periodic-solution branches by variation of a system parameter, $\lambda$, we apply the pseudo-arclength method \cite{KELLER197873, Doedel1991a, Doedel1991b, Dickson2007}. 
The solution $\mathbf{X}(0)$
and the parameter $\lambda$ are expressed as functions of a new parameter, $s$, the branch (pseudo-) arclength, i.e.: $\mathbf{X}(0) = [\mathbf{X}(0)](s)$, and $\lambda = \lambda(s)$.
By expressing the solution as a function of the arclength parameter, $\lambda$ is also an unknown and we require an additional equation, the so-called arclength constraint, which has the general form: $N(\mathbf{X}(0),\lambda)=0$. 
This results in an augmented system of non-linear equations, and the relevant linearized system at each Newton iteration is:

\begin{equation} \label{eq:NRaugm}
    \begin{bmatrix}
        \mathbf{I} - \Large{\frac{\partial \boldsymbol{\Phi}_{T_f}}{\partial \mathbf{X}(0)}} & -\Large{\frac{\partial \boldsymbol{\Phi}_{T_f}}{\partial \lambda}} \\
        \Large{\frac{\partial N}{\partial \mathbf{X}(0)}} & \Large{\frac{\partial N}{\partial \lambda}}
    \end{bmatrix} 
    \begin{bmatrix}
        \delta \mathbf{X}(0) \\
        \delta \lambda
    \end{bmatrix}
    =
    - \begin{bmatrix}
        \mathbf{X}(0) - \boldsymbol{\Phi}_{T_f} \left(\mathbf{X}(0) \right)\\
          N(\mathbf{X}(0),\lambda).
      \end{bmatrix}
\end{equation}
    
\noindent In our computations the arclength constraint has the following form:

\begin{eqnarray}
    N(\mathbf{X}(0),T_f,\lambda) =&& \left< \frac{[\mathbf{X}(0)]_1-[\mathbf{X}(0)]_0}{\delta s}, \mathbf{X}(0) - [\mathbf{X}(0)]_1 \right> + \nonumber  \\&& \frac{\lambda_1-\lambda_0}{\delta s} (\lambda - \lambda_1)-\delta s, 
\end{eqnarray}

\noindent where $\left( [\mathbf{X}(0)]_0, \lambda_0 \right)$ and $\left( [\mathbf{X}(0)]_1, \lambda_1 \right)$
respresent two previously computed periodic solutions, and $\delta s$ is the magnitude of the pseudo-arclength continuation step.

%%%%%%%%%%%%%%%%%%%%%%%%%%%%%%%%%%%%%%%%%%%%%%%%%%%%%%%%%%%%%%%%%%%%%%

\onecolumngrid

\begin{figure}[H]%[t]
    \centering
    \makebox[0pt][c]{\includegraphics[width=15cm,height=7.5cm]{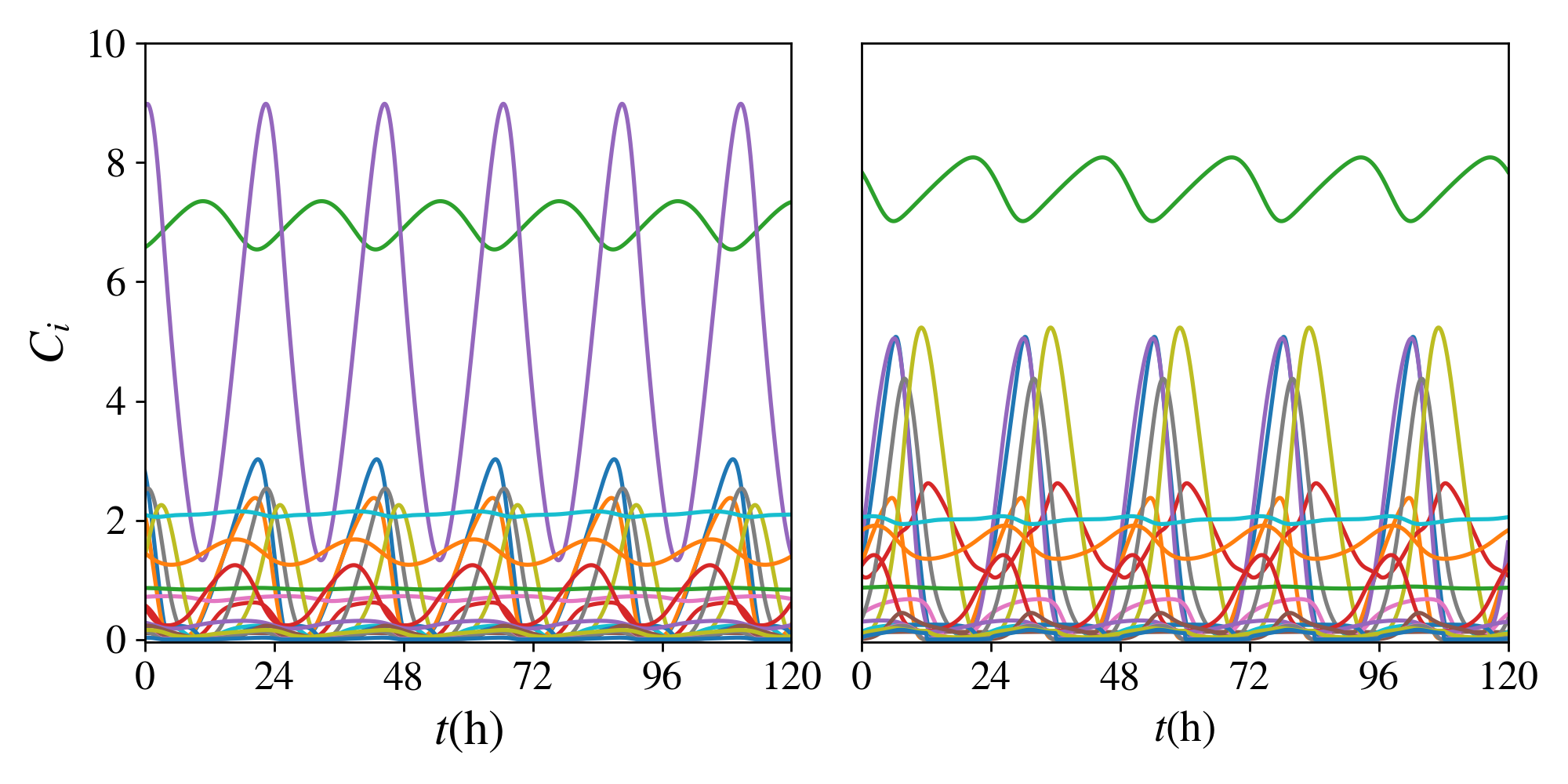}}
    \caption{Timeseries of steady state oscillations of the (left) unforced system with natural period $T_0=22.04\mathrm{h}$, and (right) forced system with $T_{f} = 24\mathrm{h}$. $C_i$ denotes the concentration of chemical species, $i$.}
    \label{Fig:Osc1}
\end{figure}
\twocolumngrid

\section{Results}
\label{Sec:results}

\subsection{Single Neuron studies}
\label{Ssec:1neuron}

When all neurons in a connected network are homogeneous, the network can in principle behave as each one of its individual neurons.  We, therefore, begin our analysis by studying the behavior of a single circadian neuron.

% \begin{center} 
%      \makebox[\textwidth][c]{\includegraphics[width=10cm,height=5cm]{Figures/Oscillation1_forced_unforced}}
%      \captionof{figure}{Timeseries of steady state oscillations of the (left) unforced system with natural period $T_0=22.04\mathrm{h}$, and (right) forced system with $T_{f} = 24\mathrm{h}$.}
% \label{Fig:Osc1}
% \end{center}

% \onecolumngrid

% \begin{figure}[H]%[t]
%     \centering
%     \makebox[0pt][c]{\includegraphics[width=15cm,height=7.5cm]{Figures/Oscillation1_forced_unforced}}
%     \caption{Timeseries of steady state oscillations of the (left) unforced system with natural period $T_0=22.04\mathrm{h}$, and (right) forced system with $T_{f} = 24\mathrm{h}$.}
%     \label{Fig:Osc1}
% \end{figure}
% \twocolumngrid

As can be seen from Fig.\ref{Fig:Osc1}, in the absence of a photic stimulus, the circadian neuron will oscillate with its intrinsic frequency, while, in the presence of a photic stimulus, the neuron can get entrained. Notice that at the frequency locked ``periodic steady state'', all the chemical species' concentrations (normalized by $1\mathrm{nM}$ for simplicity) will oscillate with the same frequency but will not necessarily be at the same phase (\textit{e.g.} not all species reach their maximum concentration simultaneously). 

\begin{figure}
    \centering
    \makebox[0pt][c]{\includegraphics[width=9cm,height=9cm]{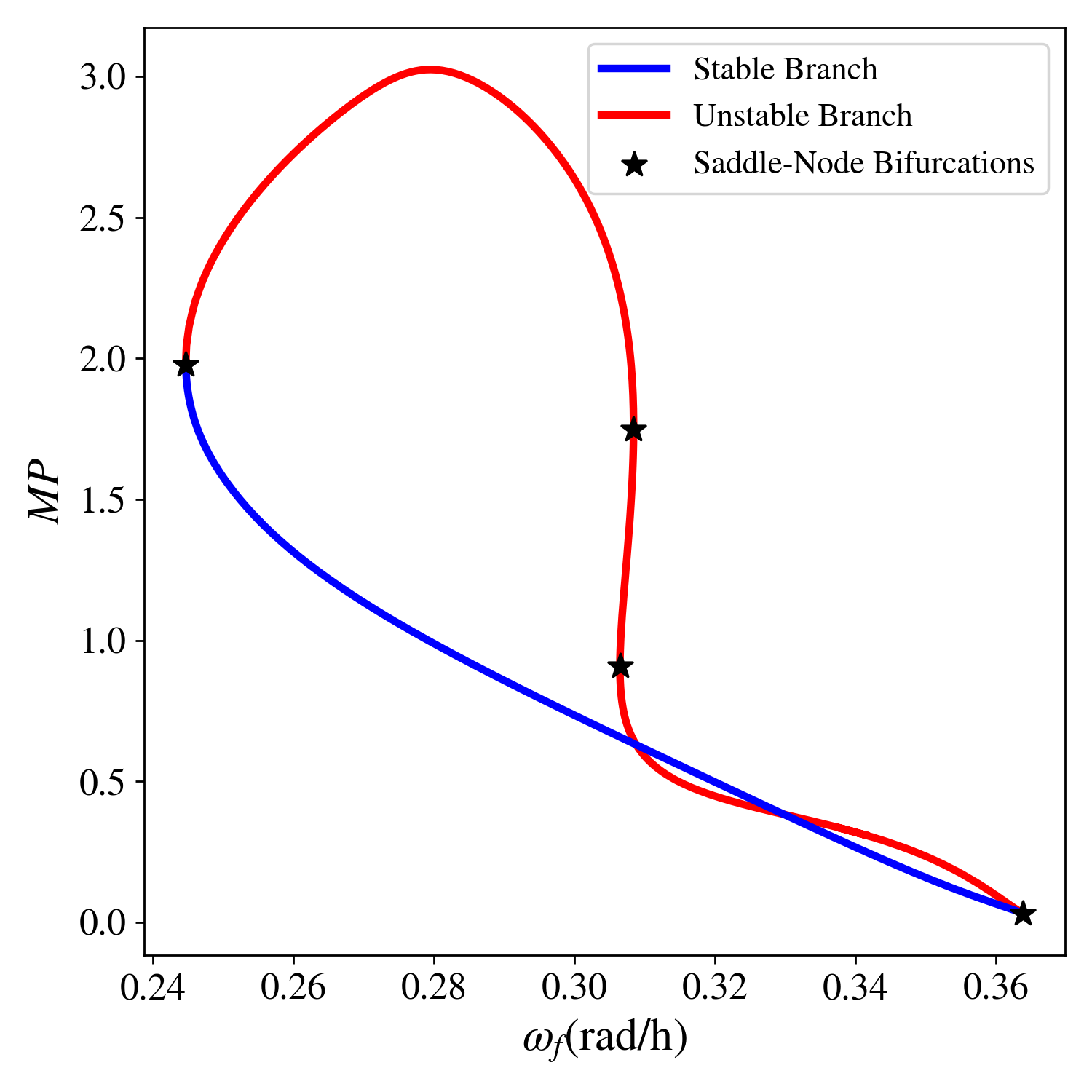}}
    \caption{Bifurcation diagram for periodic steady states under photic stimulus for one circadian neuron. Here, the \textit{MP} (\textit{Per} mRNA) projection is shown. For reference: $\omega_f = 0.262\mathrm{rad/h}$ for $T_{f}=24h$.}
    \label{Fig:Isola1}
\end{figure}

To explore entrained periodic solutions of a single circadian neuron, we perform pseudo-arclength continuation w.r.t the forcing angular frequency $\omega_f$ (reminder: $\omega_f =  \frac{2\pi}{T_{f}}$), resulting in bifurcation diagrams like the one in Fig.\ref{Fig:Isola1}.

% \begin{center} 
%      \makebox[\textwidth][c]{\includegraphics[width=9cm,height=9cm]{Figures/isola1_049omega.png}}
%      \captionof{figure}{Bifurcation diagram for periodic steady states under photic stimulus for one circadian neuron. Here, the \textit{MP} (\textit{Per} mRNA) projection is shown. For reference $\omega_f = 0.262$ for $T_{forcing}=24h$.}
% \label{Fig:Isola1}
% \end{center}

In bifurcation studies of entrainment problems, such loops are common when continuing solutions w.r.t. the forcing angular frequency \cite{Tomita1979} and are termed \textit{isolas}. Each point of the isola corresponds to a fixed point of the stroboscopic map, or, equivalently, to a periodic steady state of the original 21-dimensional ODE system. The limiting values of the bifurcation diagram w.r.t. the forcing angular frequency define {\em the limits of entrainment} and constitute saddle-node bifurcations of limit cycles. That is why the periodic solutions exchange stability at these points (see Fig. \ref{Fig:Isola1}). Along the unstable branch (red) two additional saddle-node bifurcations are observed, further destabilizing the (already unstable) entrained solution. 

\begin{figure} %3:2
    \centering
    \makebox[0pt][c]{\includegraphics[width=10.5cm,height=7cm]{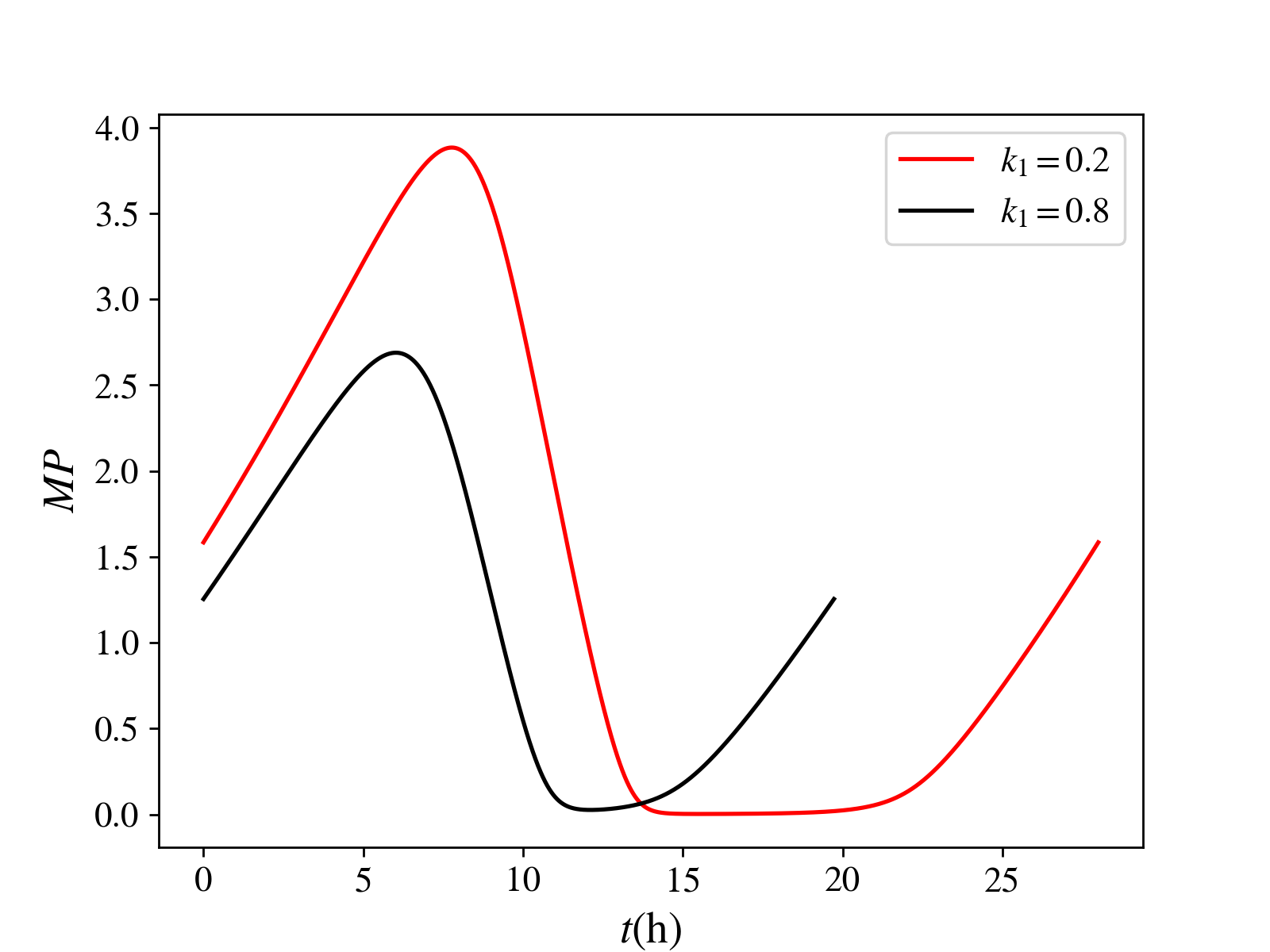}}
    \caption{Unforced \textit{MP} oscillations at the periodic steady state of $k_1=0.2$ and $k_1=0.8$.}
    \label{fig:drug_osc}
\end{figure}

To simulate the effect of the drug Longdaysin, we investigate how the synchronization limits change for different values of $k_1$ (normalized by $1\mathrm{h^{-1}}$ for simplicity). As suggested in \cite{StJohn2040} we expect decreasing values of $k_1$, (which correspond to higher doses of Longdaysin) to lead to 
longer intrinsic (unforced) periods and larger oscillation amplitudes for the non-driven neurons. First, we perform continuation of the \textit{unforced} system w.r.t. $k_1$ and
examine the limit cycles at the two extreme limits of the $k_1$
interval (Fig.\ref{fig:drug_osc}). Then, we select a discrete, representative set of $k_1 $ values, and calculate the limits of entrainment for each one of these values by constructing isolas (as in Fig. \ref{Fig:Isola1}) with respect to the forcing frequency. Note that the nominal value of $k_1$ according to \cite{VASALOU201112}  is $k_1=0.49$ (also included).

% \begin{center} 
%      \makebox[\textwidth][c]{\includegraphics[width=12cm,height=8cm]{Figures/k10208.png}}
%      \captionof{figure}{Unforced \textit{MP} oscillations at the periodic steady state of $k_1=0.2$ and $k_1=0.8$.}
% \label{fig:drug_osc}
% \end{center}

\begin{figure*} %14:17.5
    \centering
    \makebox[0pt]{\includegraphics[width=17cm,height=21.25cm]{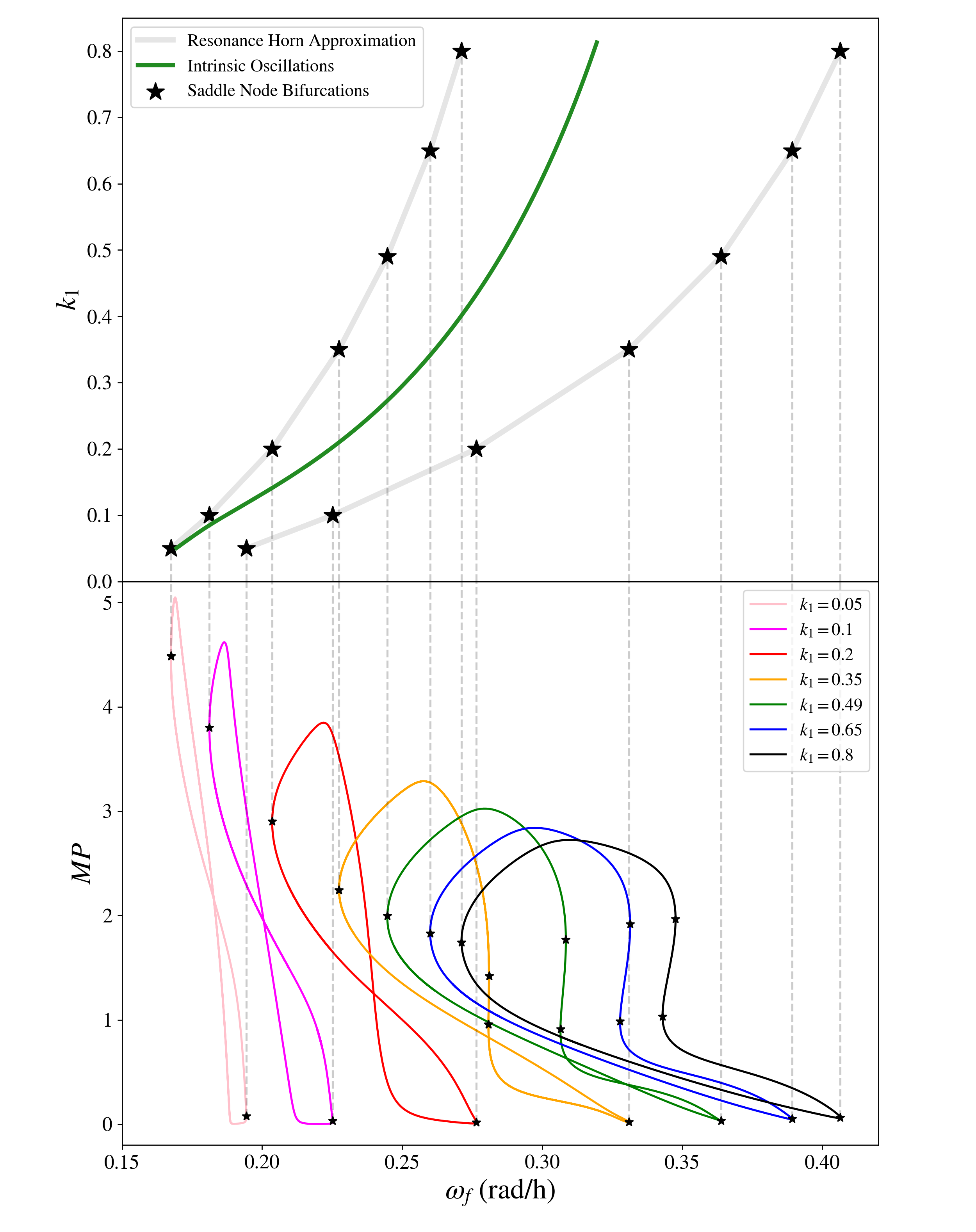}}
    \caption{Bifurcation diagram for periodic steady states under photic stimulus for varying angular frequency $\omega_f$ and Longdaysin effect ($k_1$ value) in the case of a single circadian neuron. At the bottom subfigure, a collection of isolas is shown for discrete values of $k_1$. At the top figure, the \textit{resonance horn} is approximated in the $\omega_f - k_1$ space.}
    \label{fig:Horn1}
\end{figure*}

We thus confirm that lower values of $k_1$ correspond to higher circadian periods and amplitudes of \textit{MP} oscillations. It can also be seen from Fig.\ref{fig:Horn1} that lower values of $k_1$ result in tighter entrainment regions. The ``natural'' period of unforced oscillations always lies within the respective entrainment boundaries, as discussed in \cite{Tomita1979}. The bifurcation diagram in the $k_1 - \omega_f$ space is equivalent to what is known in the literature as a \textit{resonance horn} or \textit{Arnold tongue}, and demarcates the parameter subregion where entrainment by an external forcing signal is feasible.
% \begin{center}
%      \makebox[\textwidth][c]{\includegraphics[width=16cm,height=20cm]{Figures/horn1.png}}
%      \captionof{figure}{Bifurcation diagram for periodic steady states under photic stimulus for varying angular frequency $\omega_f$ and Longdaysin effect ($k_1$ value) in the case of a single circadian neuron. At the bottom subfigure, a collection of isolas is shown for discrete values of $k_1$. At the top figure, the \textit{resonance horn} is approximated in the $\omega_f - k_1$ space.}
% \label{fig:Horn1}
% \end{center}

% \begin{center}
%     \includegraphics[width=12cm, height=10cm]{Figures/neuron_bifurcations}
% \end{center}

% \begin{center}
%     \includegraphics[width=12cm, height=6cm]{Figures/insets.png}
% \end{center}

\begin{figure*}
     \centering
     \begin{subfigure}[b]{0.45\textwidth}
         \centering
         \includegraphics[width=\textwidth]{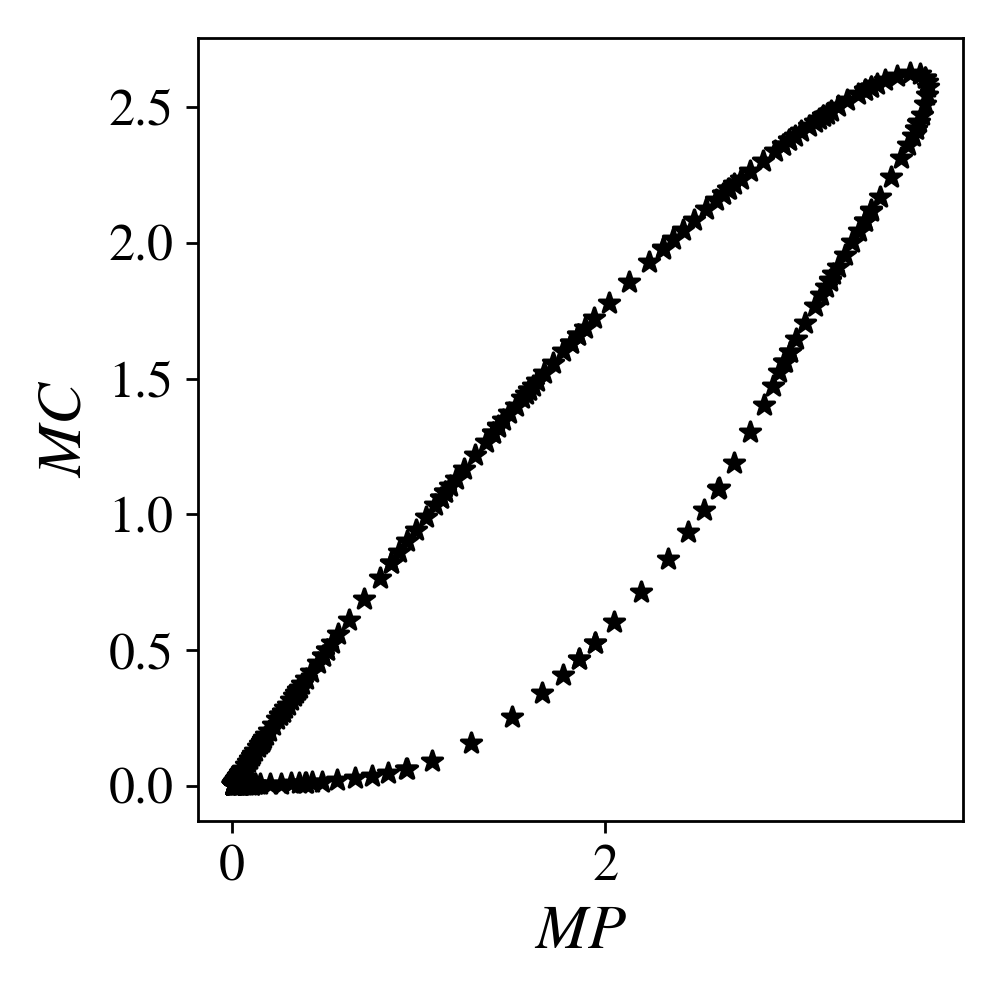}
        %  \caption{$k_1=0.05, \omega_f = 0.167$}
         \label{fig:quasiperiodic_torus1}
     \end{subfigure}
     \hfill
     \begin{subfigure}[b]{0.53\textwidth}
         \centering
         \includegraphics[width=\textwidth]{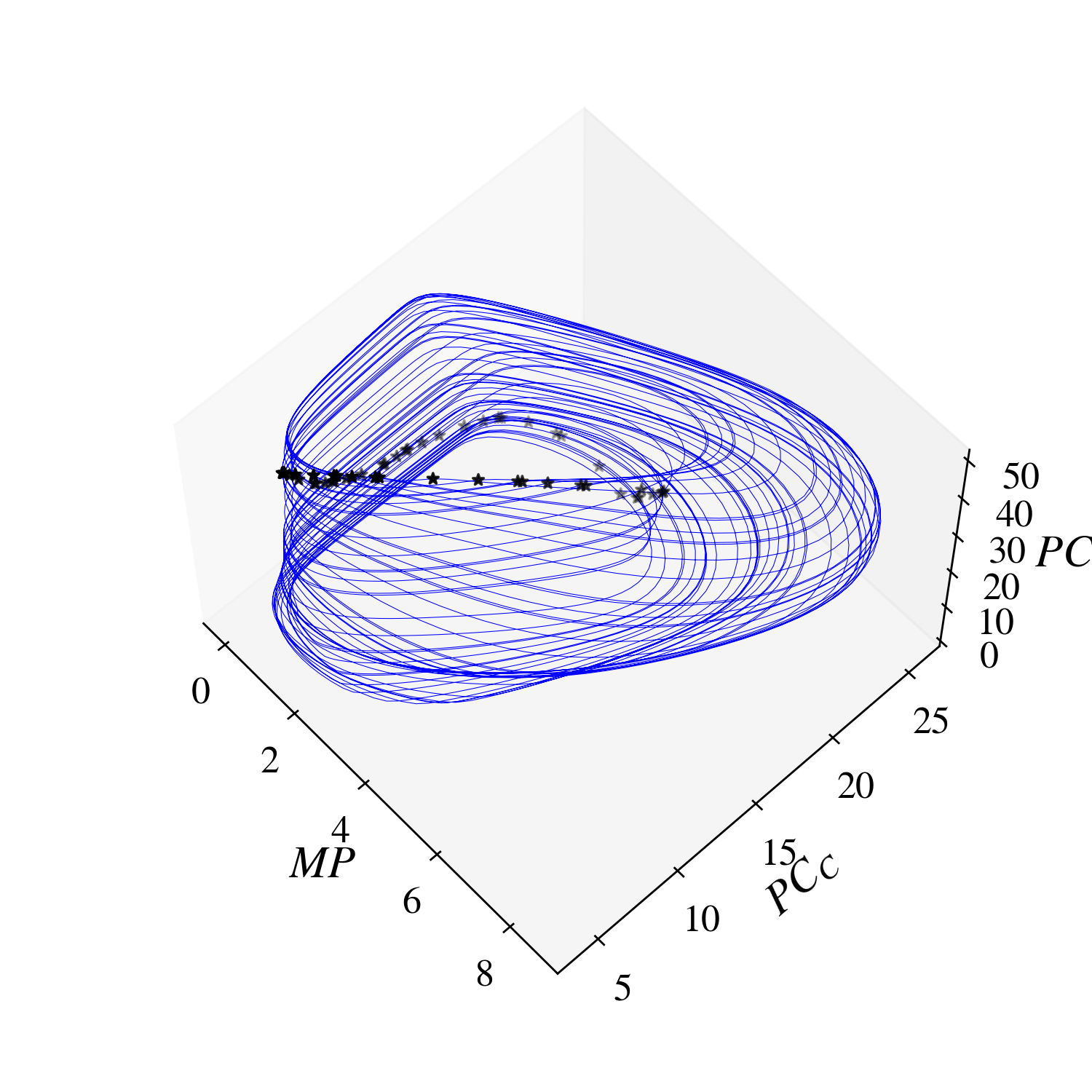}
        %  \caption{$\phi = 0.085$}
         \label{fig:quasiperiodic_torus2}
     \end{subfigure}
     \caption {Loss of entrainment for $k_1=0.05, \omega_f = 0.16746 \mathrm{rad/h}$. Near the \textit{tip} of the resonance horn (low $k_1$), quasiperiodicity is observed outside, but close to the entrainment limits. This can be confirmed by plotting iterates of the stroboscopic map (here, shown in the $MC-MP$ projection), where an invariant circle is observed (left). In the phase portrait representation (right), trajectories are attracted to a torus (here, the $MP-PC_C-PC$ projection is shown, along with stroboscopic map iterates). Note that $MP$: \textit{Per} mRNA, $MB$: \textit{Bmal1} mRNA, $MC$: \textit{Cry} mRNA, $PC$: nonphosphorylated \textit{Cry} protein in the cytosol,  $PC_C$: nonphosphorylated \textit{Per-Cry} protein complex in the cytosol.}
     \label{fig:quasiperiodic}
\end{figure*}

Outside of the resonance horn in the $k_1 - \omega_f$ space, various other dynamic responses are expected, \textit{e.g.} quasiperiodicity, frequency locking, or chaos \cite{Tomita1979}. Fig. \ref{fig:quasiperiodic} shows quasiperiodic response for low values of $k_1$, at $\omega_f=0.16746 \mathrm{rad/h}$ right after crossing the left saddle-node bifurcation (at $\omega_f=0.16749 \mathrm{rad/h}$). Indeed, as it can be seen in the right panel of Fig. \ref{fig:quasiperiodic} the trajectories are now attracted to a high-dimensional torus.

\begin{figure*}
     \centering
     \begin{subfigure}[b]{0.49\textwidth}
         \centering
         \includegraphics[width=\textwidth]{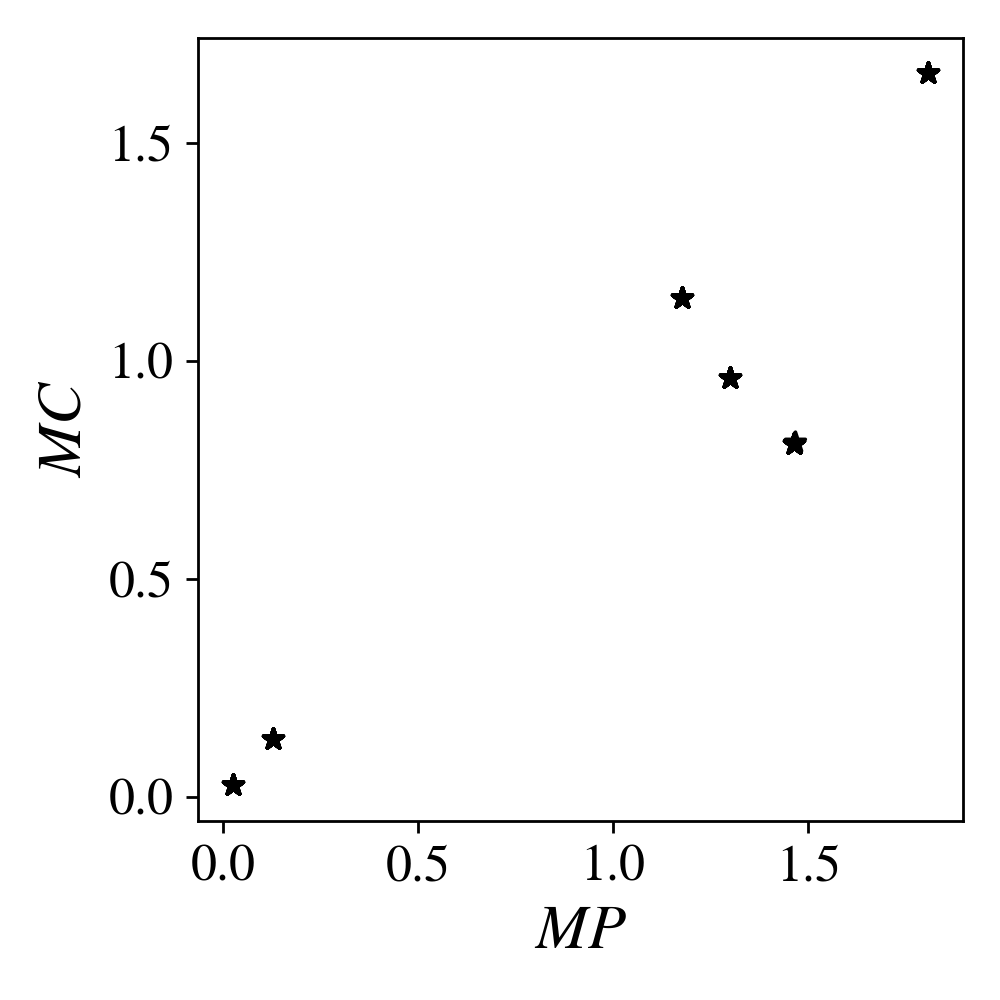}
        %  \caption{$k_1=0.05, \omega_f = 0.167$}
         \label{fig:chaos1}
     \end{subfigure}
     \hfill
     \begin{subfigure}[b]{0.49\textwidth}
         \centering
         \includegraphics[width=\textwidth]{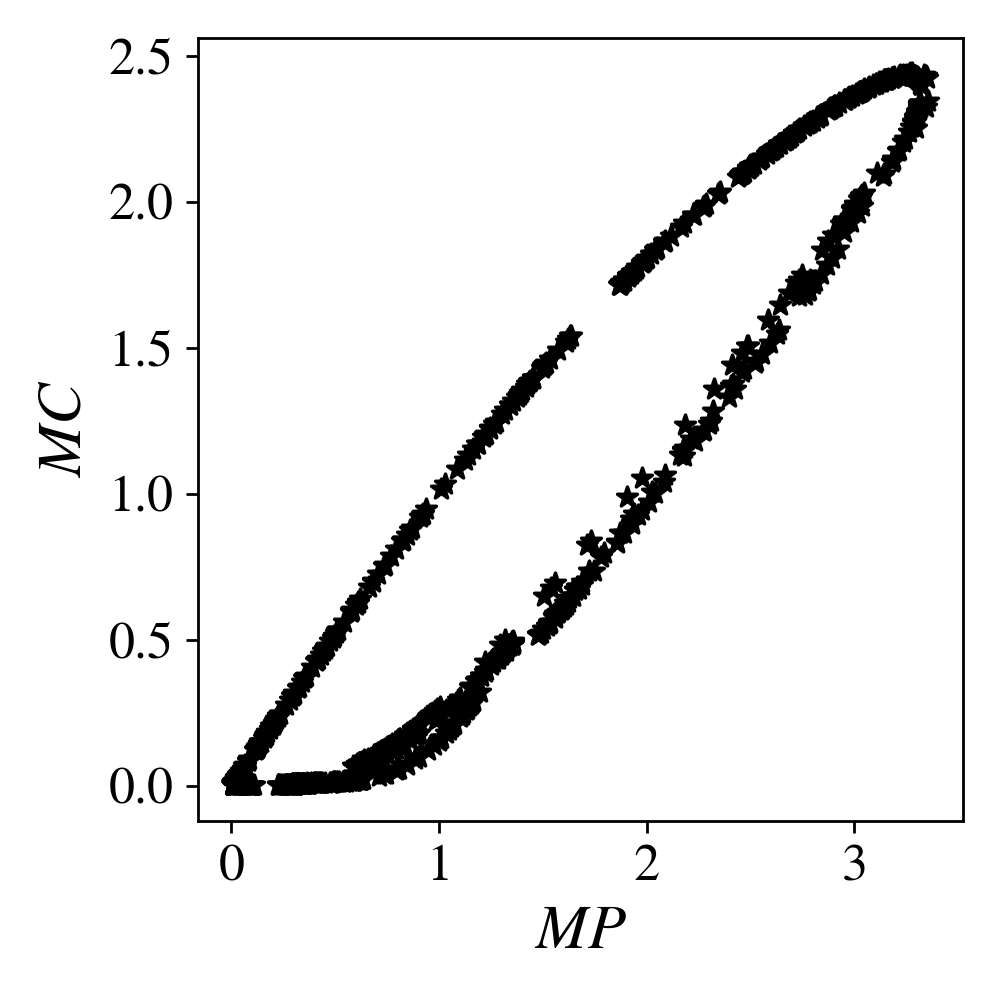}
        %  \caption{$\phi = 0.085$}
         \label{fig:chaos2}
     \end{subfigure}
     \caption {Loss of entrainment for $k_1=0.49: \omega_f = 0.2446 \mathrm{rad/h}$. (left panel) $\omega_f = 0.3639 \mathrm{rad/h}$ (right panel). Stroboscopic map iterates reveal frequency-locking and chaos respectively (here in the $MC-MP$ projection)}
     \label{fig:high_k1}
\end{figure*}

For higher $k_1$ values it is expected that different phenomena arise \cite{KEVREKIDIS19861263, Tomita1979}. Indeed, for $k_1=0.49$ and $\omega_f=0.2446 \mathrm{rad/h}$ a period-6 solution arises (frequency locking) after crossing the left bifurcation at $\omega_f=0.2447 \mathrm{rad/h}$ (left panel, Fig. \ref{fig:high_k1}). Crossing the other limit of entrainment for $k_1=0.49$ (saddle-node bifurcation at $\omega_f=0.3638 \mathrm{rad/h}$), chaos is observed to emerge for $\omega_f=0.3639 \mathrm{rad/h}$ (right panel, Fig. \ref{fig:high_k1}).

Figs.\ref{fig:quasiperiodic}, \ref{fig:high_k1} stand as evidence that periodically forced circadian neurons demonstrate the entire gamut of dynamic responses typical of periodically forced dynamical systems \cite{Tomita1979}.

Subsequently, we perform continuaton of the periodic steady states w.r.t the duty cycle ($\phi$) for a \textit{fixed} period (here, $T_{f} = 24h$ or, equivalently, $\omega_f = 0.268 \mathrm{rad/h}$). In other words, we investigate entrainment of circadian neurons when the ratio of the length of the day (and, correspondingly, the duration of the night) changes.

\begin{figure}[H]%16.5:11
    \centering
    \makebox[0pt]{\includegraphics[width=16.5cm,height=11cm]{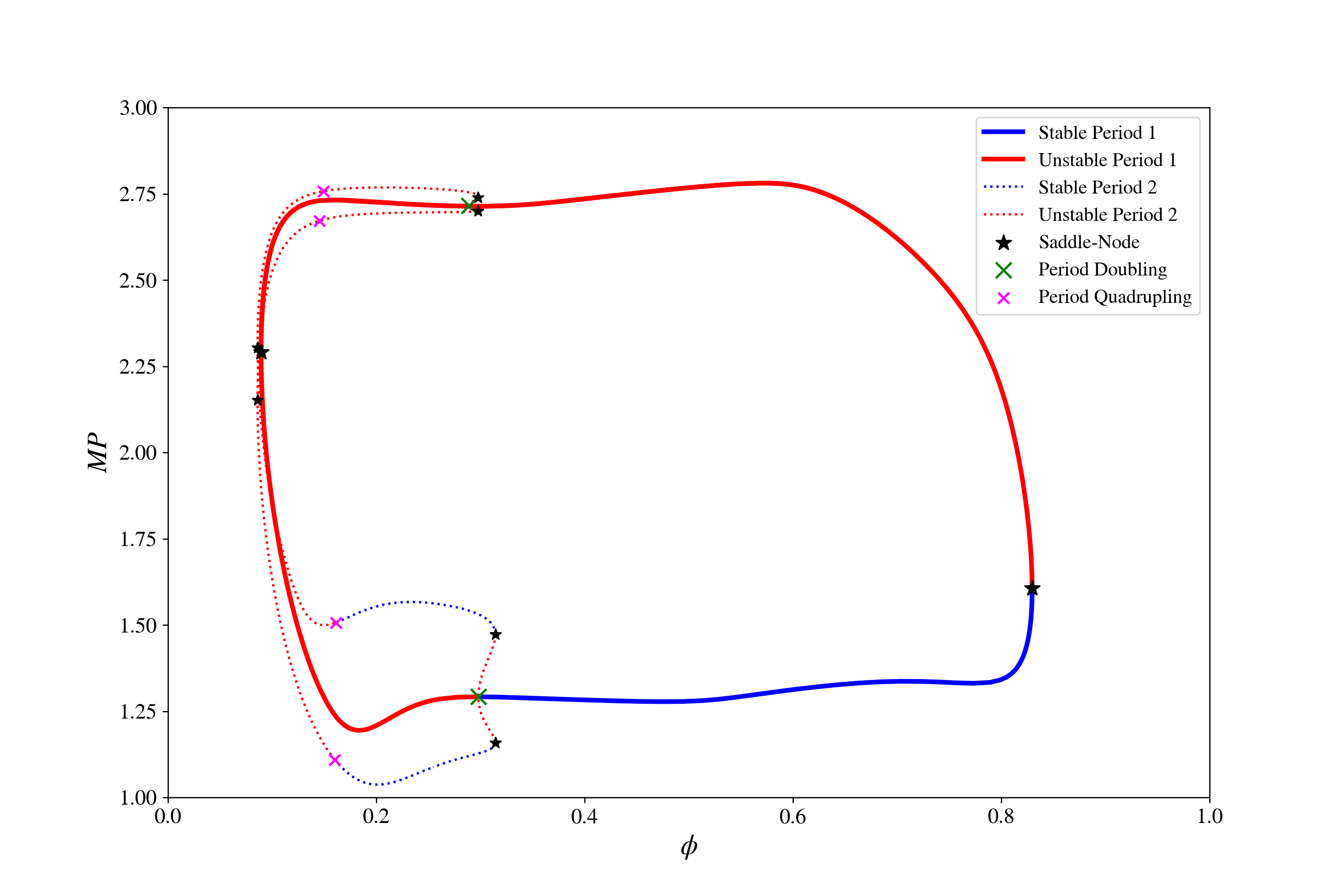}}
    \caption{Bifurcation diagram for periodic steady states w.r.t the duty cycle $\phi$ under photic stimulus of fixed angular frequency $\omega_f = 0.268 \mathrm{rad/h}$ and without drug intervention ($k_1 = 0.49$) in the case of a single circadian neuron. Notice the additional bifurcations (leading to complex dynamics) towards lower day/night ratios.}
    \label{fig:phi1}
\end{figure}

As seen in Fig.\ref{fig:phi1} there are entrainment limits w.r.t to the duty cycle as well. As anticipated, circadian neurons are unable to synchronize when the day/night phases become highly unbalanced. However, entrainment is not always lost in exactly the same way as the typical isolas we have encountered thus far. 

\begin{figure}[H]
     \centering
     \begin{subfigure}[b]{0.49\textwidth}
         \centering
         \includegraphics[width=\textwidth]{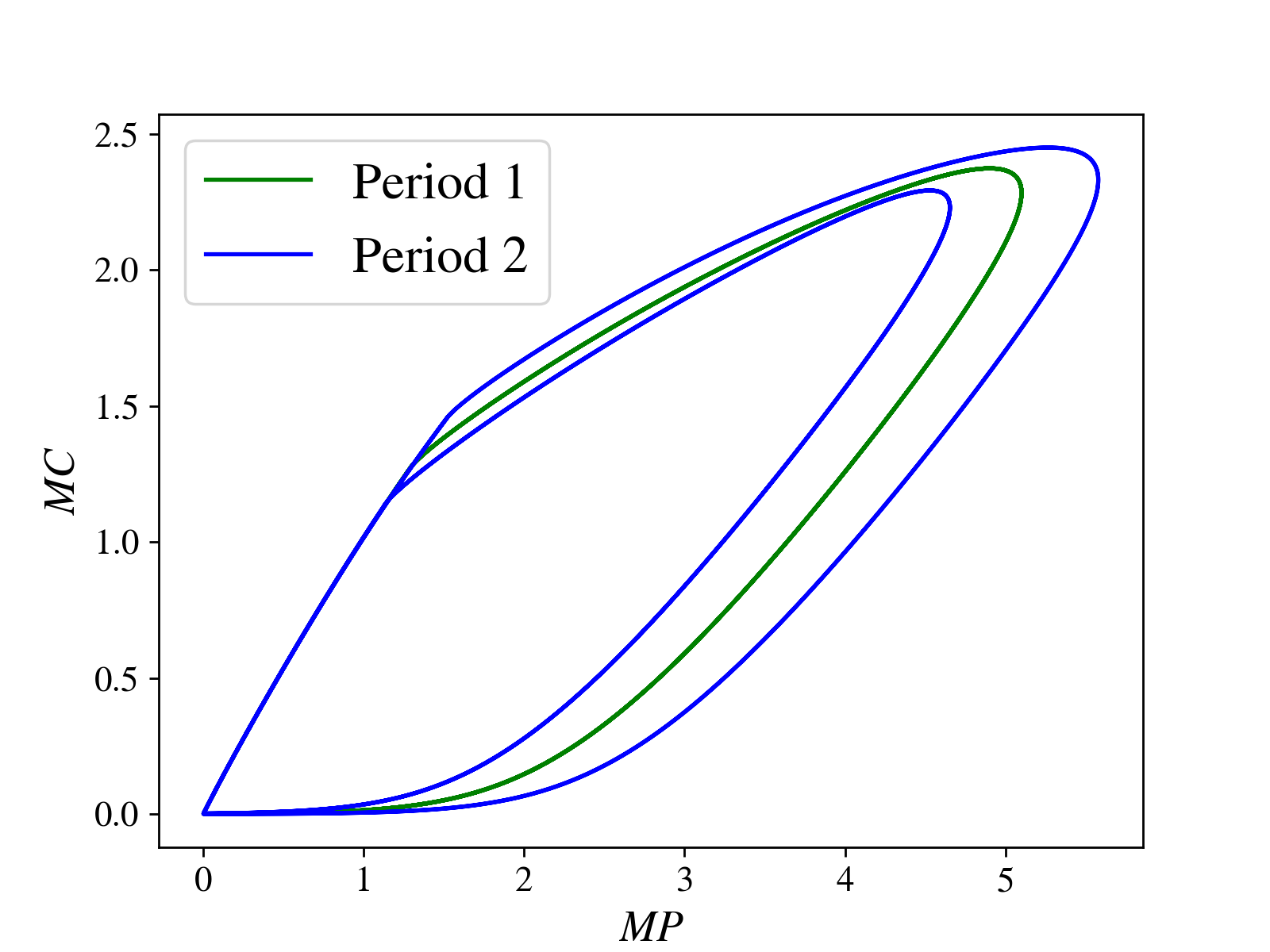}
         \caption{$\phi = 0.305$}
         \label{fig:bist1}
     \end{subfigure}
     \hfill
     \begin{subfigure}[b]{0.49\textwidth}
         \centering
         \includegraphics[width=\textwidth]{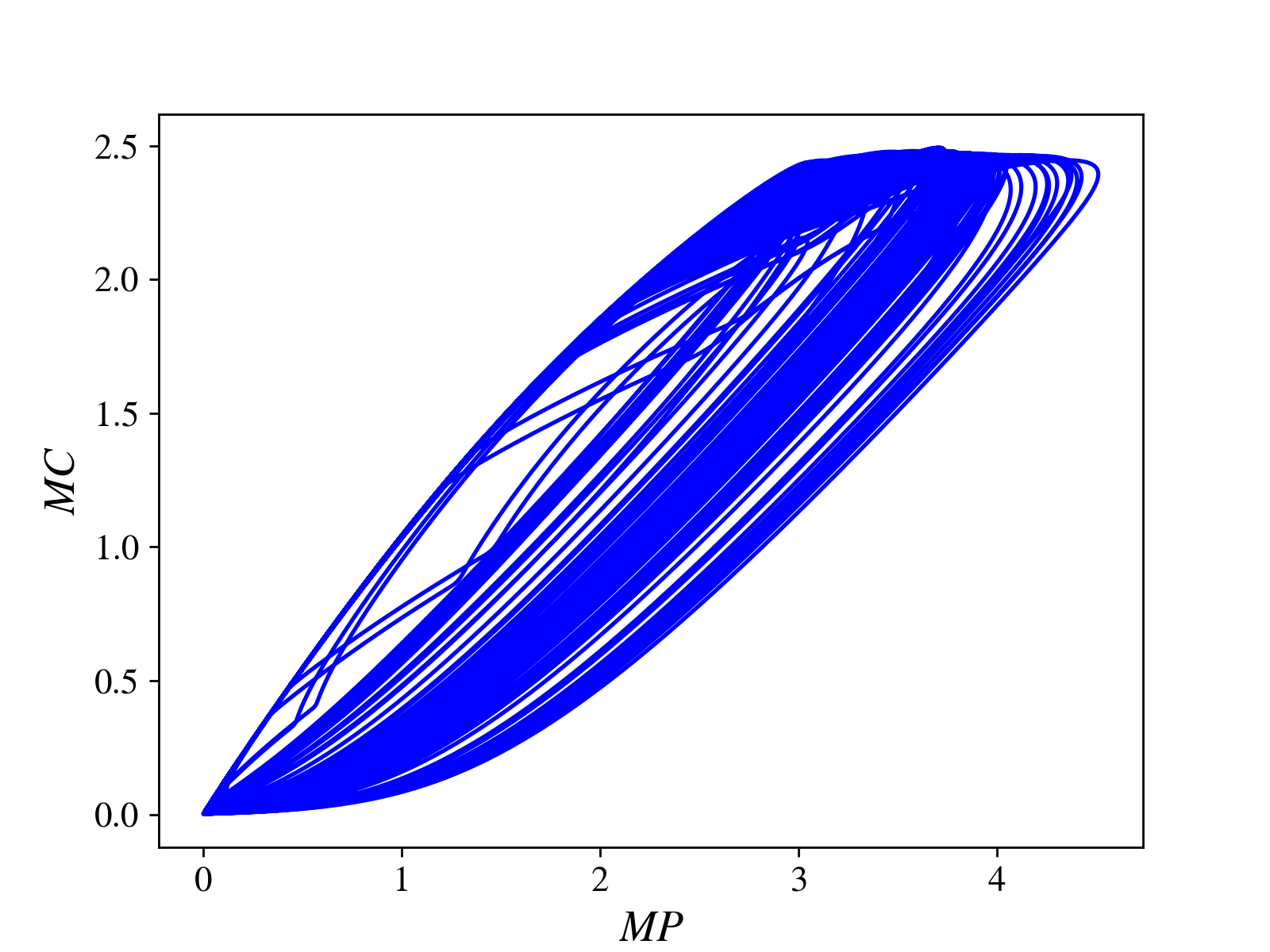}
         \caption{$\phi = 0.085$}
         \label{fig:chaos}
     \end{subfigure}
     \caption {Phase portraits at low $\phi$ values where bistability as well as chaos is observed.}
     \label{fig:special_phi}
\end{figure}

%as in Fig. \ref{Fig:Isola1}: 
For high day/night phase ratios, a saddle-node bifurcation marks the upper boundary of entrainment;  but at low day/night ratios, the system undergoes a period doubling bifurcation (both the stable and the unstable branch). This means that the circadian neurons will return to the same state after two forcing periods. Interestingly, for some region, stable period-1 and period-2 solutions coexist, which means that circadian neurons are briefly bi-stable (Fig.\ref{fig:bist1}). It is worth noting that the period-2 branches undergo an additional period-doubling bifurcation, leading to period-4 solutions. We hypothesize that this is the beginning of a period-doubling route to chaos (Fig. \ref{fig:chaos}) through a cascade of period-doubling bifurcations.

Note that integration of the ODEs describing neuronal dynamics of a single neuron was performed in Python's 
{\fontfamily{qcr}\selectfont
solve\_ivp}
platform, using the default 
{\fontfamily{qcr}\selectfont
RK23
}
integrator (Explicit Runge-Kutta method of order 3(2) \cite{BOGACKI1989321}), with absolute and relative tolerance at $10^{-8}$. Results where confirmed against MATLAB using the {\fontfamily{qcr}\selectfont
ode23} solver with relative tolerance at $10^{-7}$. Newton-Krylov GMRES was performed with our Python own code with tolerance $10^{-5}$, implemented as in \cite{Kelley2004}. Pseudo-arclength continuation was performed with our own Python code similarly to \cite{Doedel1991a}, with an adaptive step, $\delta s$ (in the range $10^{-3} -10^{-1}$). The numerics for the neuronal network, results of which are presented in Section  \ref{Ssec:network} are in similar ranges.

\onecolumngrid

\twocolumngrid

\onecolumngrid

\twocolumngrid

\vspace{0cm}
\subsection{Heterogeneity in the  Neuronal Network} 
\label{Ssec:network}

In a realistic neuronal network each neuron is expected to be unique, or, in terms of computational modeling, to have its own intrinsic parameter values. As described in Section \ref{Sec:model}, here we consider networks of 425 neurons that are heterogeneous w.r.t. the values of the parameter $v_{sP_0}$, for each neuron.

The effect of heterogeneity can be seen when plotting projections of the (now, $8925-$dimensional) limit cycle for two variables of each  neuron (Fig. \ref{fig:lc_all}). We plot these limit cycles for two extents of heterogeneity (variance of the -here, normal- distribution of the heterogeneous parameter) and for five different simulated Longdaysin effects ($k_1$). As seen in Figs. \ref{fig:lc_all} and \ref{fig:high_lc} increasing heterogeneity causes the trajectories of different neurons to move further apart. However, all neurons follow qualitatively similar trajectories  since the neuronal network is entrained.

\onecolumngrid

\begin{figure}[H]

\makebox[\linewidth]{
\begin{subfigure}{.5\textwidth}
  \centering
  % include first image
  \includegraphics[width=0.95 \linewidth]{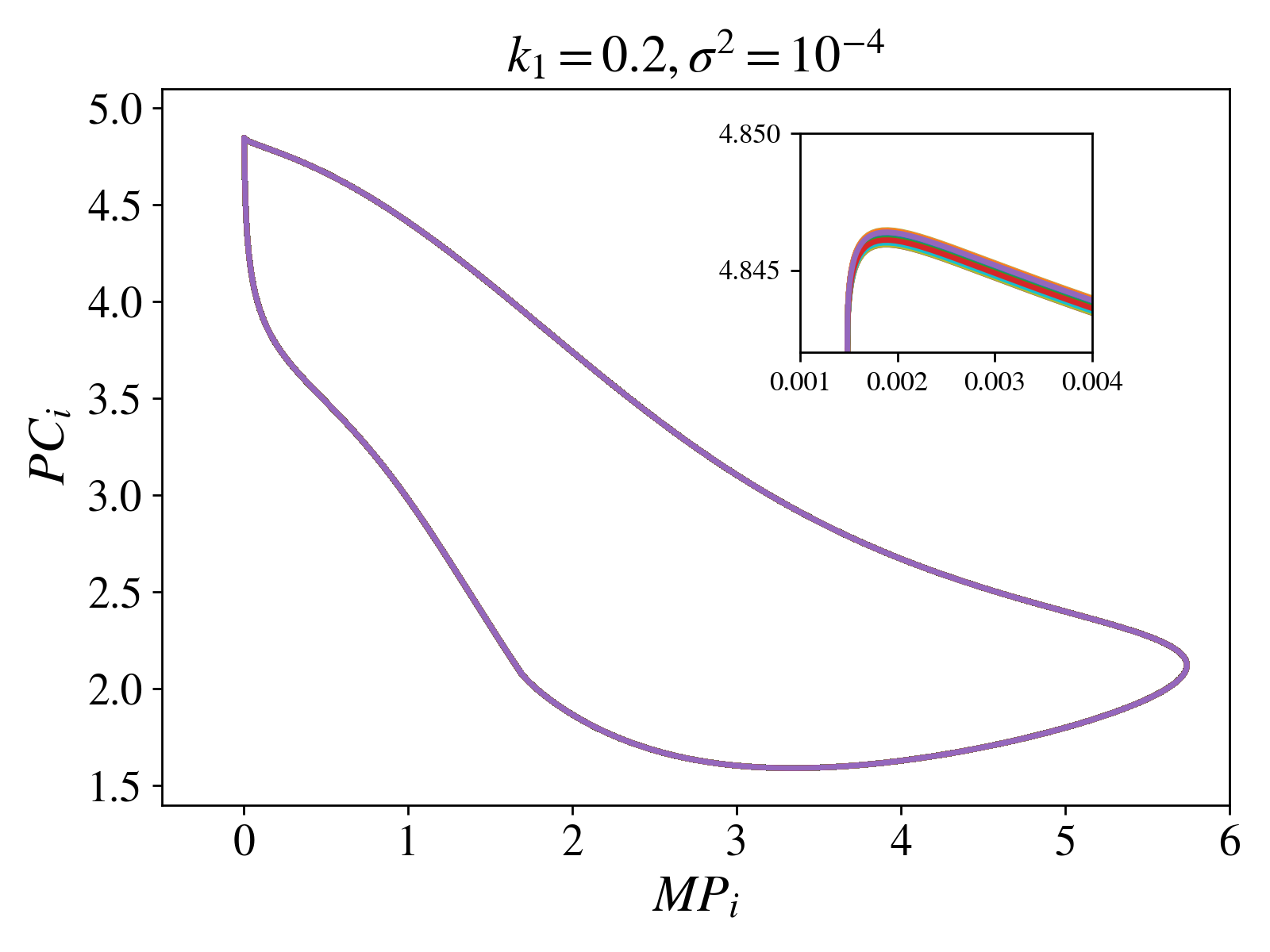} 
  \caption{}
  \label{fig:sub-first}
\end{subfigure}
\begin{subfigure}{.5\textwidth}
  \centering
  % include second image
  \makebox[0pt]{\includegraphics[width=0.95\linewidth]{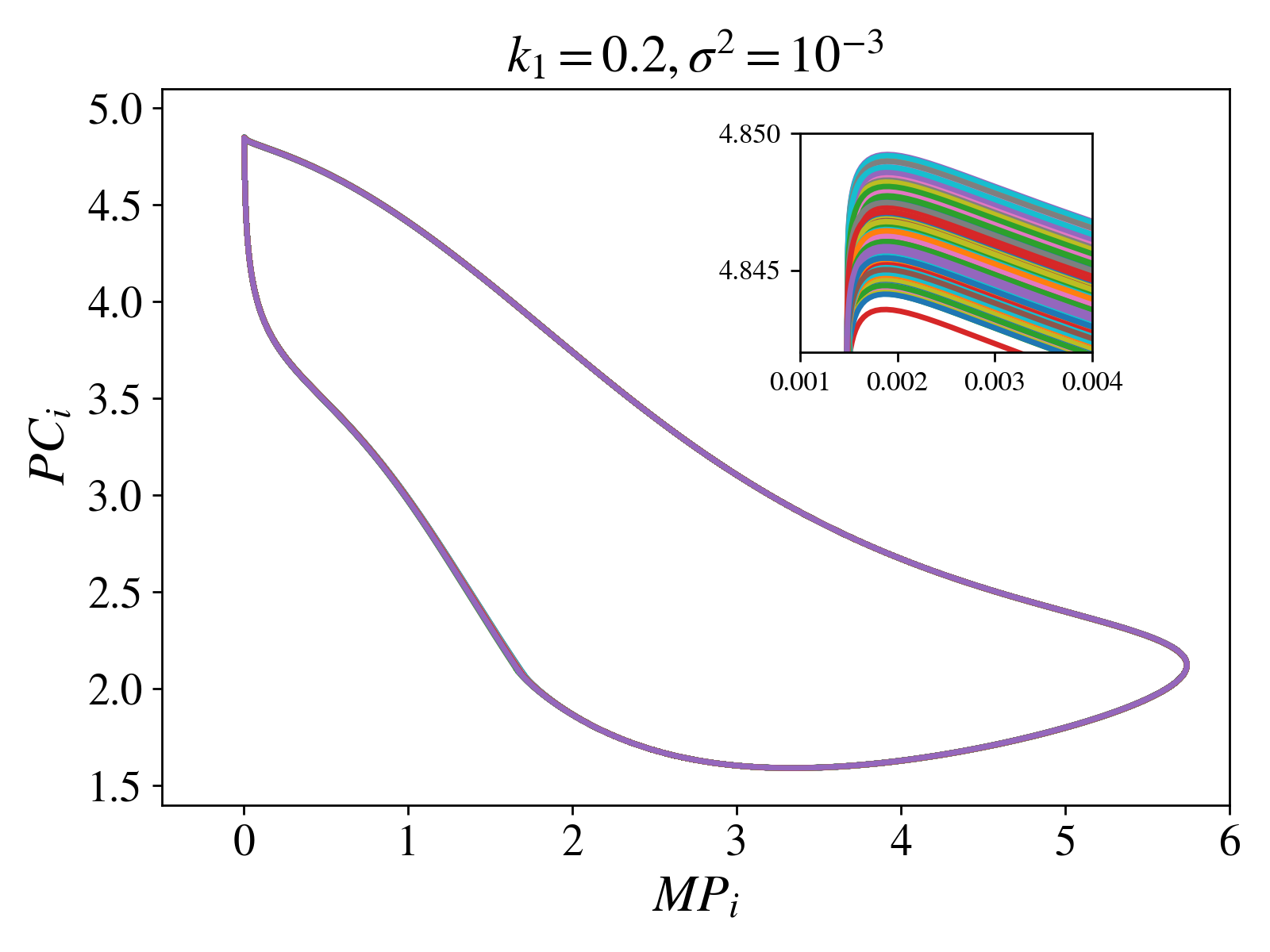}}  
  \caption{}
  \label{fig:sub-second}
\end{subfigure}}
\newline

\makebox[\linewidth]{\begin{subfigure}{.5\textwidth}
  \centering
  % include third image
  \includegraphics[width=0.95\linewidth]{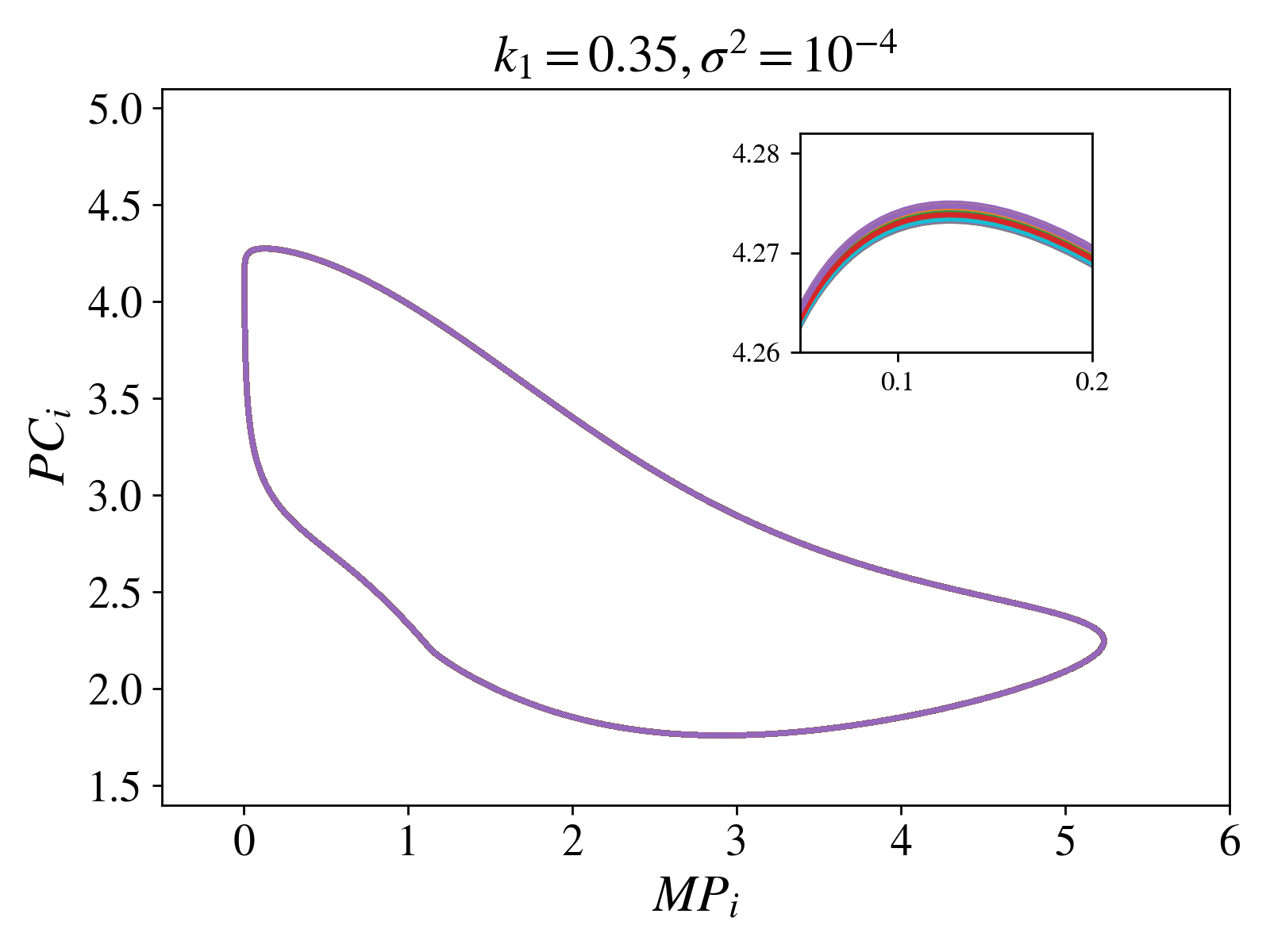}  
  \caption{}
  \label{fig:sub-third}
\end{subfigure}
\begin{subfigure}{.5\textwidth}
  \centering
  % include fourth image
  \includegraphics[width=0.95\linewidth]{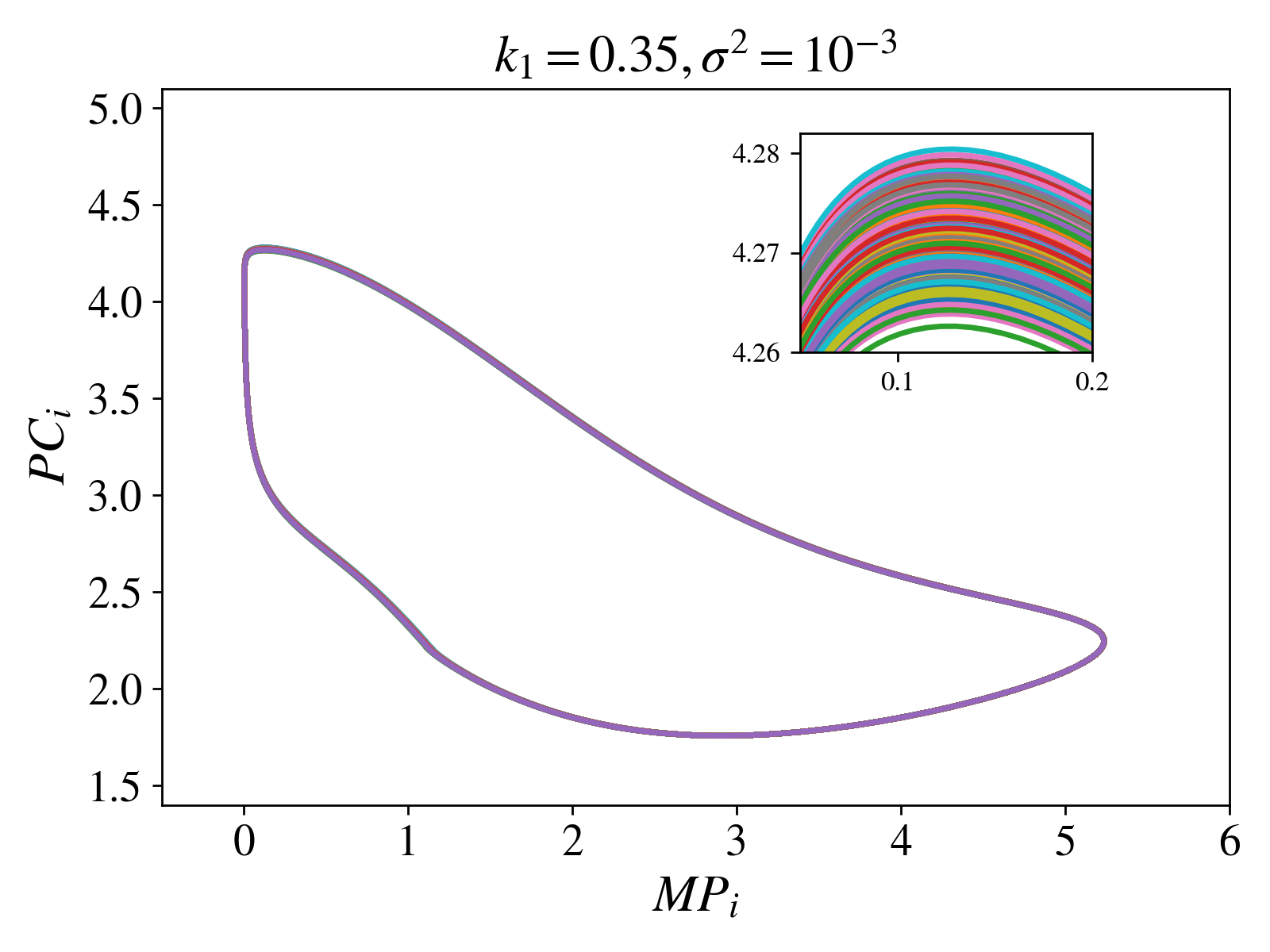}  
  \caption{}
  \label{fig:sub-fourth}
\end{subfigure}}

\makebox[\linewidth]{\begin{subfigure}{.5\textwidth}
  \centering
  % include third image
  \includegraphics[width=0.95\linewidth]{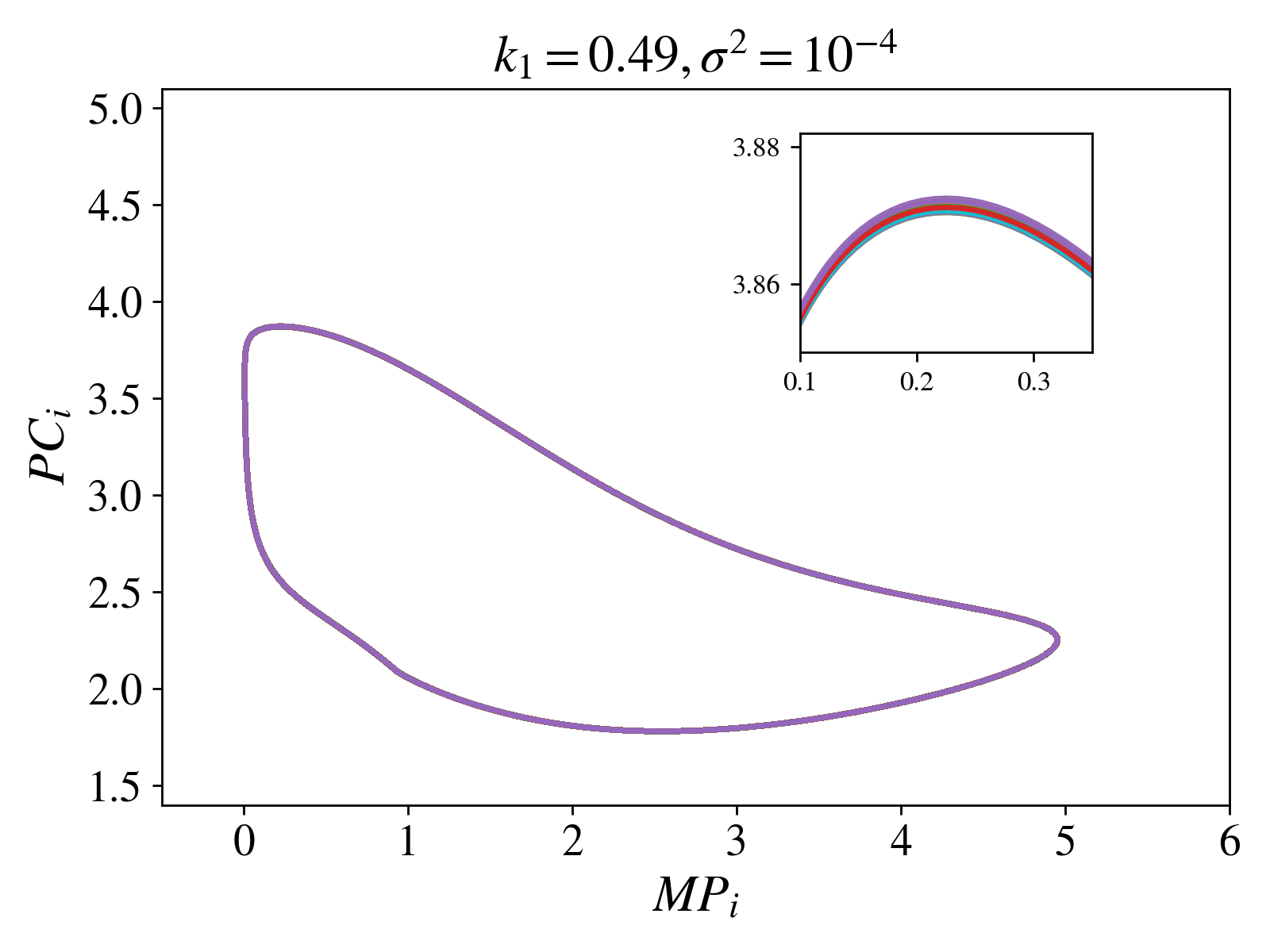}  
  \caption{}
  \label{fig:sub-fifth}
\end{subfigure}
\begin{subfigure}{.5\textwidth}
  \centering
  % include fourth image
  \includegraphics[width=0.95\linewidth]{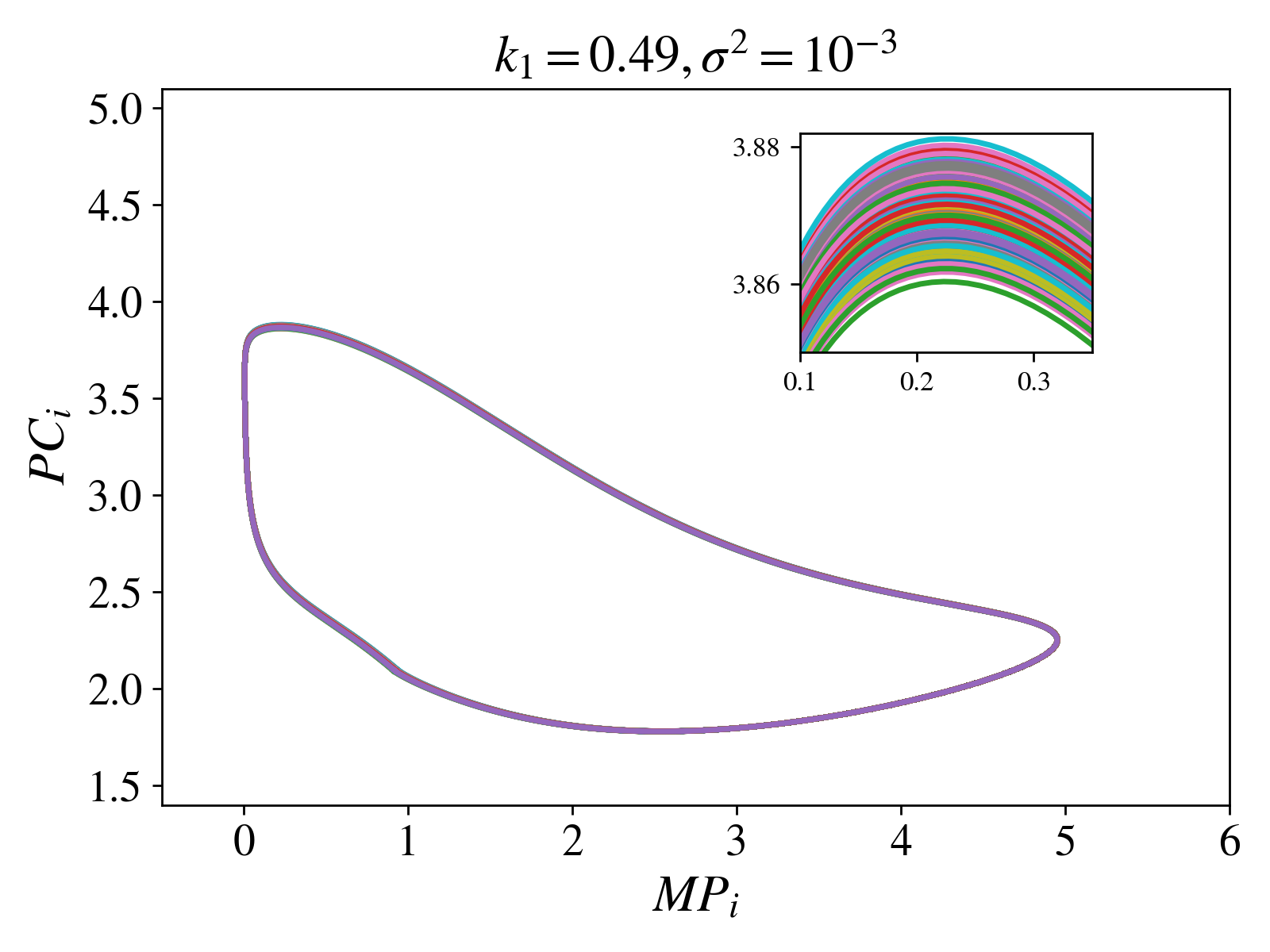}  
  \caption{}
  \label{fig:sub-sixth}
\end{subfigure}}

\label{fig:fig}
\end{figure}
% \newpage
\begin{figure}[H]\ContinuedFloat
\makebox[\linewidth]{\begin{subfigure}{.5\textwidth}
  \centering
  % include third image
  \includegraphics[width=1.\linewidth]{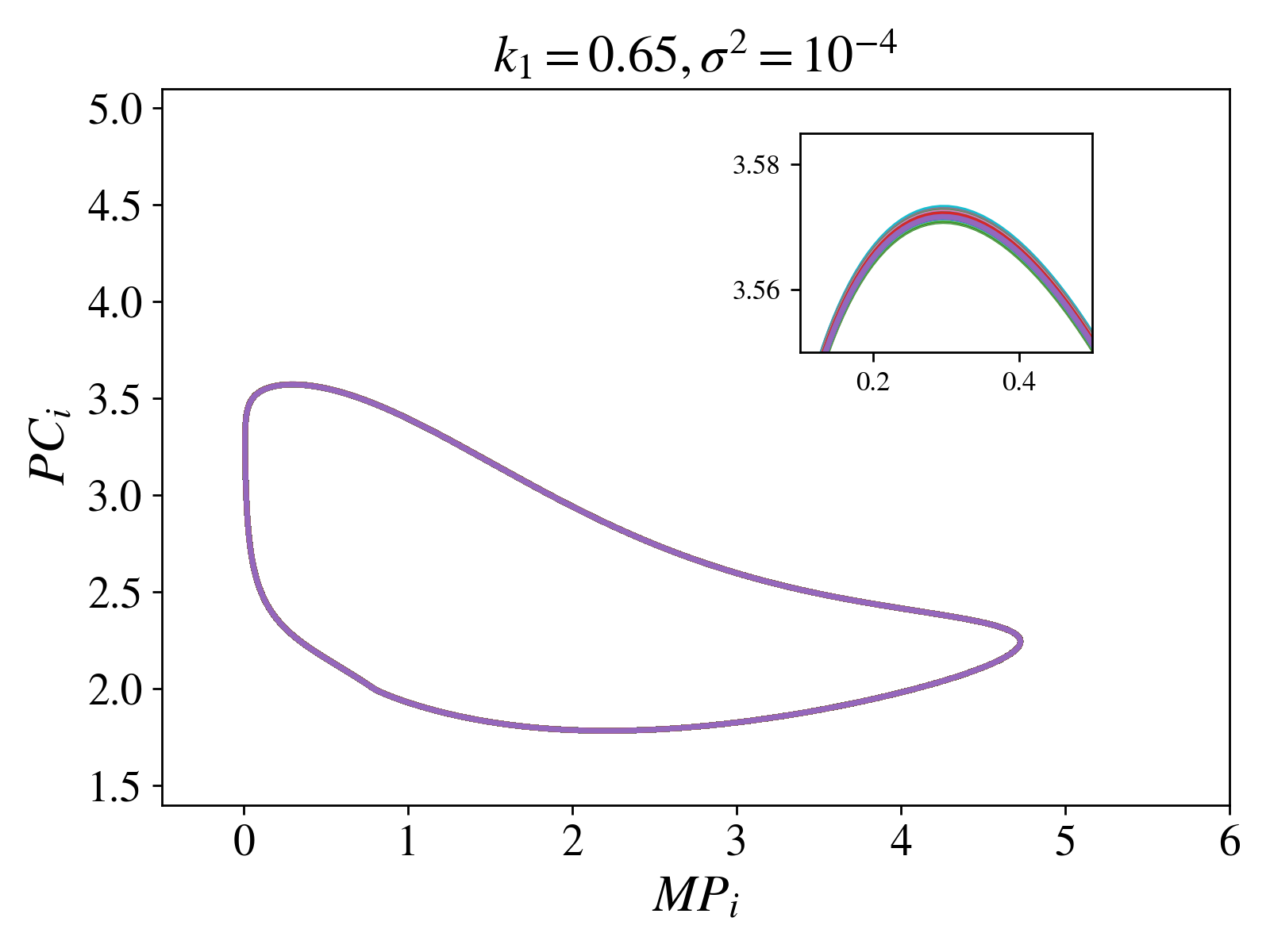}  
%   \figuretag{(g)}
  \caption{} 
  \label{fig:sub-seventh} 
\end{subfigure}
\begin{subfigure}{.5\textwidth}
  \centering
  % include fourth image
  \includegraphics[width=1.\linewidth]{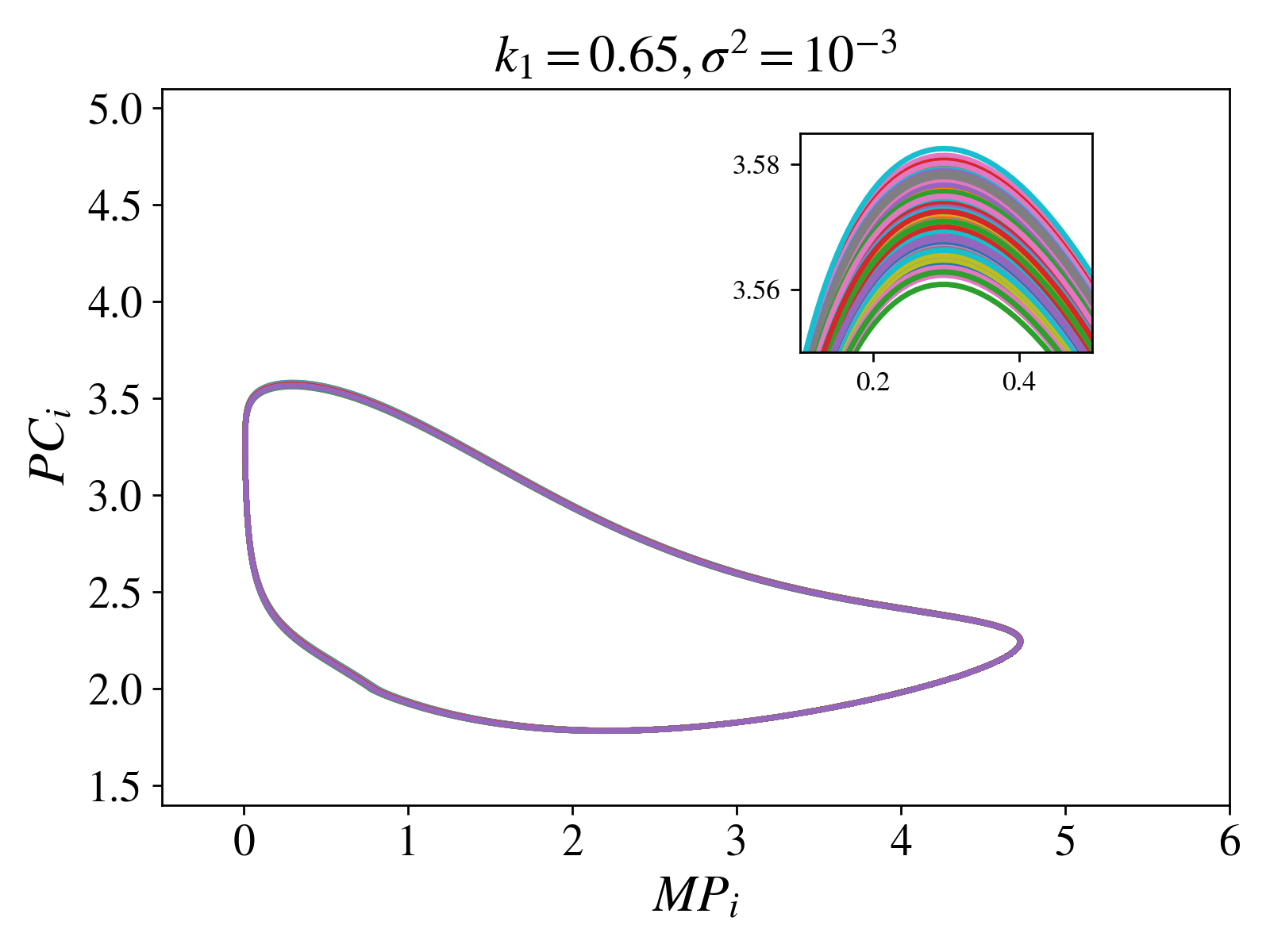}  
  \caption{}
  \label{fig:sub-eighth}
\end{subfigure}}

\makebox[\linewidth]{\begin{subfigure}{.5\textwidth}
  \centering
  % include third image
  \includegraphics[width=1.\linewidth]{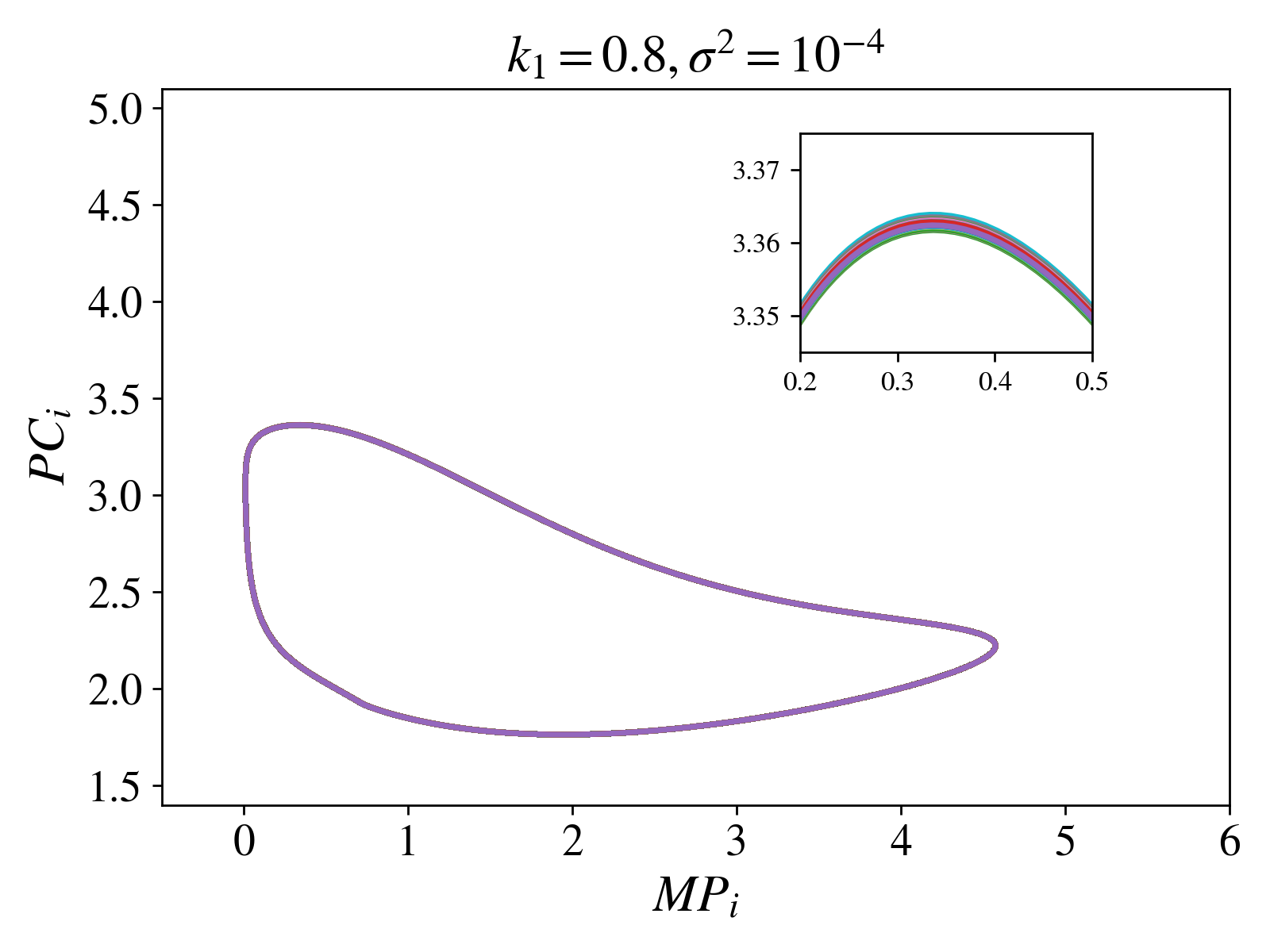}  
  \caption{}
  \label{fig:sub-ninth}
\end{subfigure}
\begin{subfigure}{.5\textwidth}
  \centering
  % include fourth image
  \includegraphics[width=1.\linewidth]{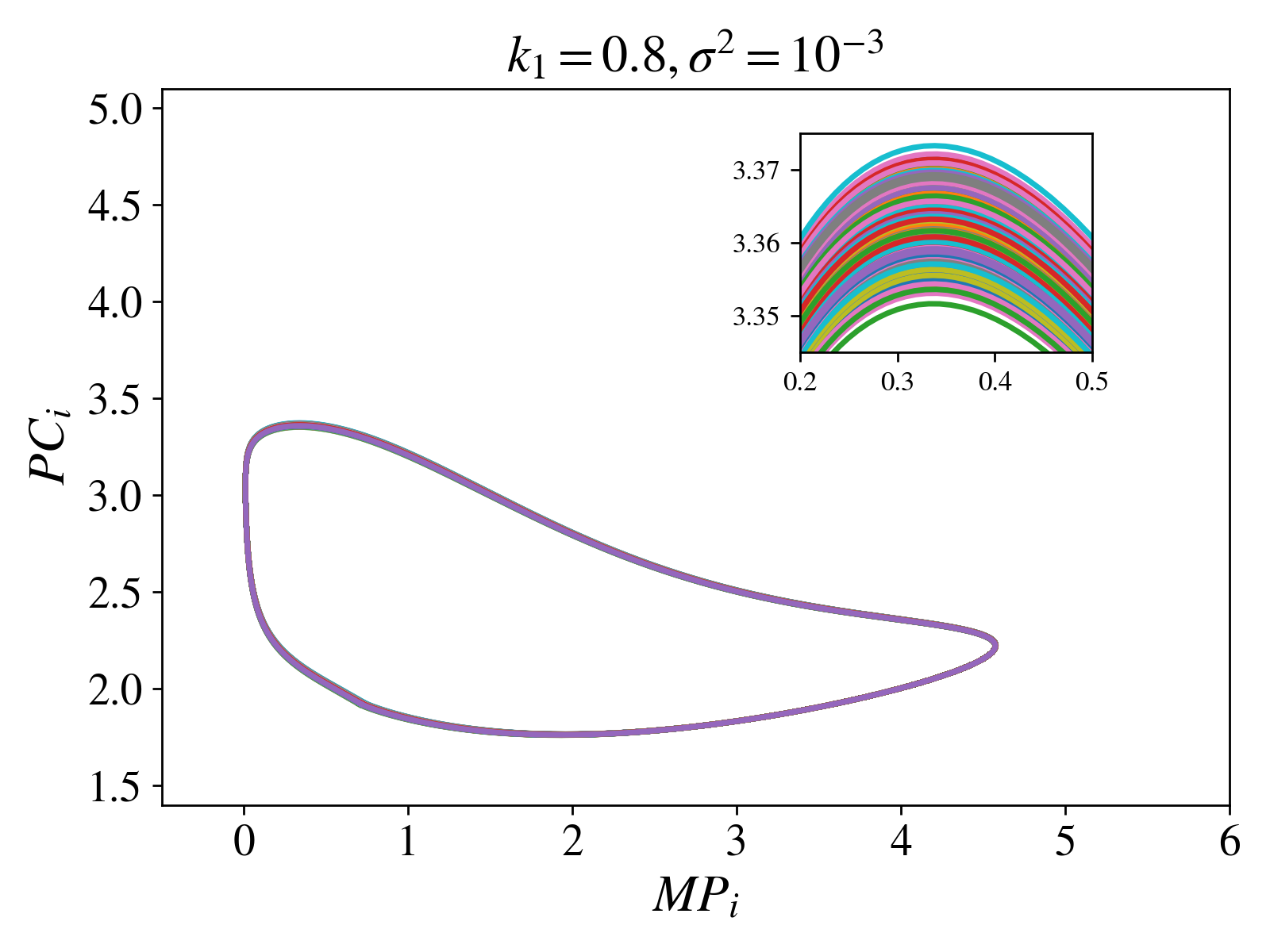}  
  \caption{}
  \label{fig:sub-tenth}
\end{subfigure}}

\caption{2D projections of the $8925-$dimensional limit cycles ($i=1,...,8925)$), for $k_1$ in $\{0.2, 0.35, 0.49$ (nominal value)$, 0.65, 0.8\}$, for heterogeneity variance $10^{-3}$ or $10^{-4}$, with forcing angular frequency equal to the intrinsic one (for each $k_1$) and for $\phi=0.5$. Here the variables $PC_i$ and $MP_i$ are reported for every neuron, $i$.}
\label{fig:lc_all}
\end{figure}
\twocolumngrid

% \begin{center} 
%      \makebox[\textwidth][c]{\includegraphics[width=12cm,height=8cm]{Figures/LC_high_049.png}}
%      \captionof{figure}{2D projections of the $8925-$dimensional limit cycle for an even higher value of heterogeneity variance studied here ($\sigma^2=4 \cdot 10^{-2}$), for comparison purposes. This limit cycle was calculated for $k_1=0.49, \phi=0.5$ with forcing period $T_f=24\mathrm{h}$. This figure should be qualitatively compared to  Figs. \ref{fig:sub-third}, \ref{fig:sub-fourth}.}
% \label{fig:high_lc}
% \end{center}

\onecolumngrid

\begin{figure}[H] %12:8
    \centering
    \makebox[0pt]{\includegraphics[width=12cm,height=8cm]{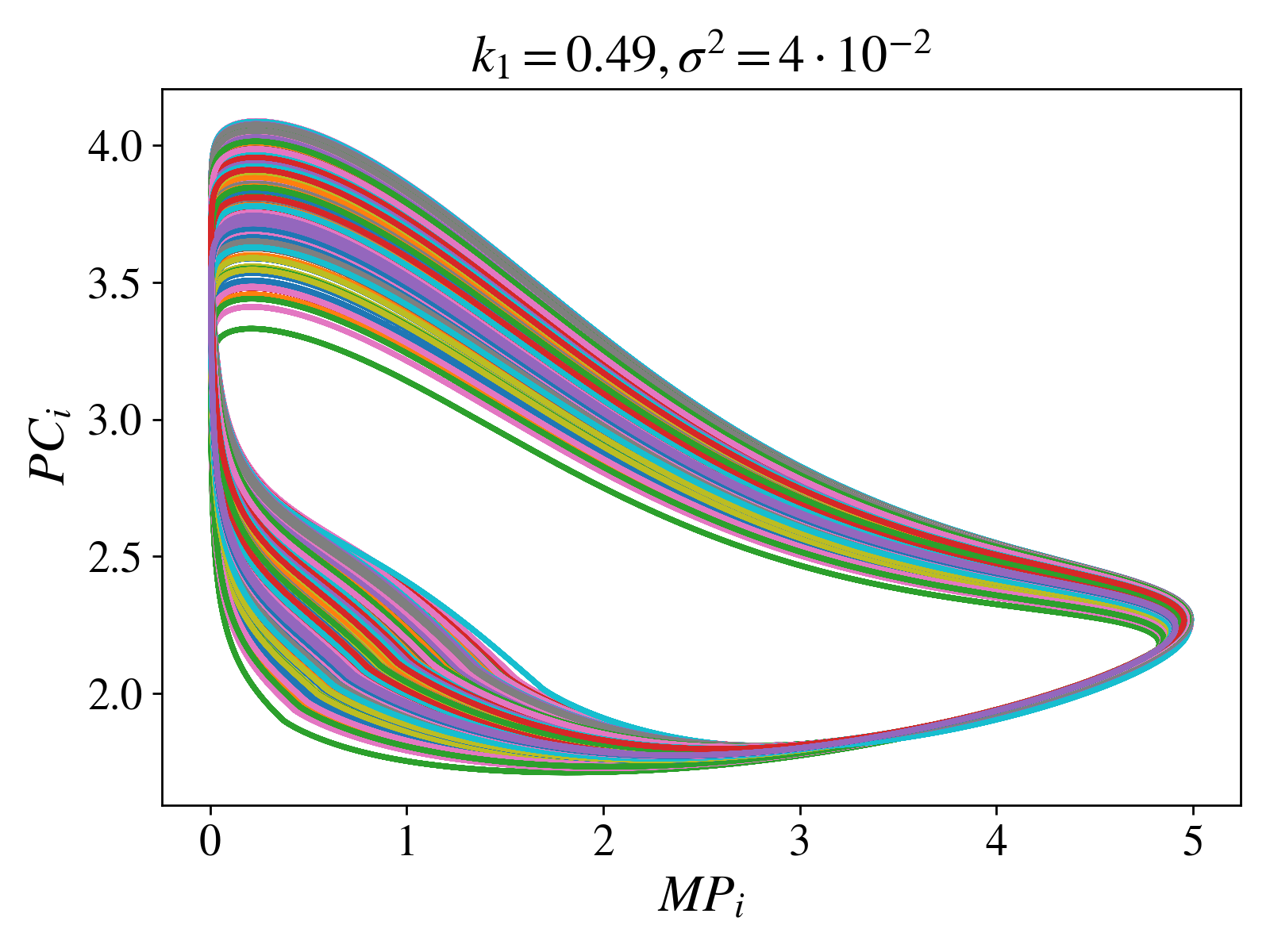}}    
    \caption{2D projections of the $8925-$dimensional limit cycle for an even higher value of heterogeneity variance studied here ($\sigma^2=4 \cdot 10^{-2}$), for comparison purposes. Here the variables $PC_i$ and $MP_i$ are reported for every neuron, $i$. This limit cycle was calculated for $k_1=0.49, \phi=0.5$ with forcing period $T_f=24\mathrm{h}$. This figure should be qualitatively compared to  Figs. \ref{fig:sub-fifth}, \ref{fig:sub-sixth}.}
    \label{fig:high_lc}
\end{figure}
\twocolumngrid

\onecolumngrid

\begin{figure}[H] %14:6
    \centering
    \makebox[0pt]{\includegraphics[width=14cm,height=6cm]{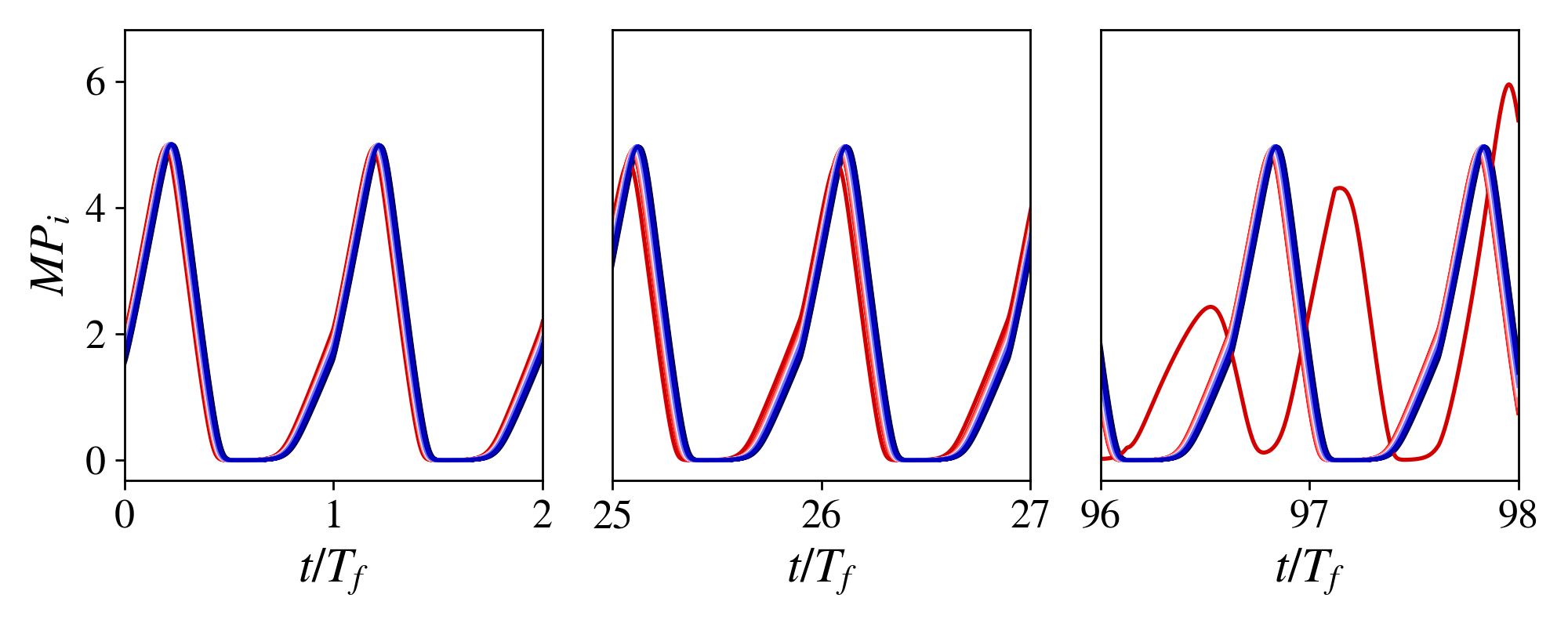}}
    \caption{Loss of entrainment due to the emergence of a single ``rogue'' oscillator. Three panels show oscillation ``snapshots'' along a long trajectory. Initially (left panel), all neurons seem to be oscillating in synchrony. After some time (middle panel), the trajectory of one neuron slowly diverges from the rest, and for longer times (right panel) it starts oscillating erratically. Variable $MP_i$ is reported for every neuron $i$, while each neuron is colored by its heterogeneity $v_{sP_0}$ value; the rogue oscillator has the highest. Forcing parameters and heterogeneity variance value are mentioned in the text.}
    \label{fig:sniper}
\end{figure}
\twocolumngrid

By analogy with Fig.\ref{fig:Horn1}, we can computationally construct a resonance horn for each heterogeneity extent (Fig. \ref{fig:hornN}). The entrainment limits do not change much with increasing neuron heterogeneity. This would seem to suggest robustness of the neuronal network to heterogeneity variations \cite{Komin2011-ob}. 

% \begin{center} 
%      \makebox[\textwidth][c]{\includegraphics[width=12cm,height=12cm]{Figures/horn1.png}}
%      \captionof{figure}{Superimposed resonance horns for the case of a single circadian neuron, and a neuronal network with varying extents of heterogeneity.}
% \label{fig:hornN}
% \end{center}

\begin{figure*}
    \centering
    \includegraphics[width=12cm,height=12cm]{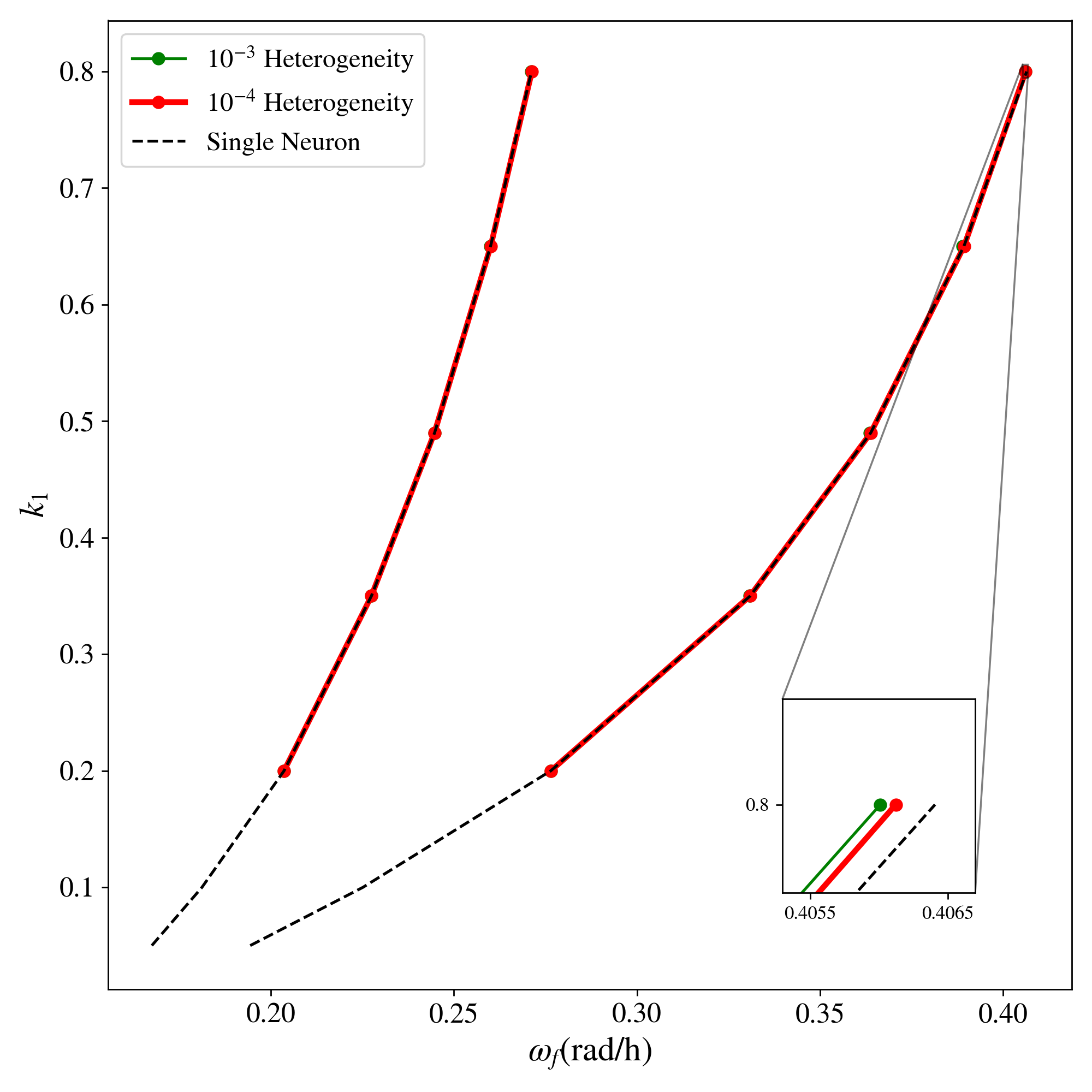}
    \caption{Superimposed resonance horns for the case of a single circadian neuron, and neuronal networks with varying extents of heterogeneity.}
    \label{fig:hornN}
\end{figure*}

As in the case of single neuron studies, we can explore how entrainment is lost in the neuronal network. As can be seen in Fig. \ref{fig:sniper}, after crossing the right saddle-node bifurcation for ($k_1=0.49$ and $\sigma^2=10^{-2}, \omega_f=0.2455 \mathrm{rad/h}$) a single ``rogue'' oscillator emerges.  By that, we mean that even though most neurons appear to oscillate in synchrony, one of them gradually desynchronizes and starts oscillating ``on its own'' with varying amplitude. The emergence of such a ``rogue'' oscillator can be attributed to large variance of the heterogeneity parameter and is associated with bifurcations of the autonomous dynamical system \cite{Bold2007}. In the dynamical systems literature a simple caricature of such a bifurcation is provided by the SNIPER (saddle-node infinite period) bifurcation \cite{Bertalan2017, Moon2006}.

% \begin{center} 
%      \makebox[\textwidth][c]{\includegraphics[width=14cm,height=6cm]{Figures/sniper.png}}
%      \captionof{figure}{Loss of entrainment due to the emergence of a single ``rogue'' oscillator. Three panels, show oscillation ``snapshots'' along a long trajectory. Initially (left panel), all neurons seem to be oscillating in synchrony. After some time (middle panel), the trajectory of one neuron slowly diverges from the rest, and for longer times (right panel) it starts oscillating erratically.}
% \label{fig:sniper}
% \end{center}

% \onecolumngrid

% \begin{figure}[H] %14:6
%     \centering
%     \makebox[0pt]{\includegraphics[width=14cm,height=6cm]{Figures/sniper.png}}
%     \caption{Loss of entrainment due to the emergence of a single ``rogue'' oscillator. Three panels, show oscillation ``snapshots'' along a long trajectory. Initially (left panel), all neurons seem to be oscillating in synchrony. After some time (middle panel), the trajectory of one neuron slowly diverges from the rest, and for longer times (right panel) it starts oscillating erratically.}
%     \label{fig:sniper}
% \end{figure}
% \twocolumngrid

Similarly to Fig.\ref{fig:phi1} we can also continue periodic solutions of the high dimensional, heterogeneous neuronal network w.r.t. $\phi$, the day/night ratio (Fig.\ref{fig:phiN}). As expected, the bifurcation diagram for a single circadian neuron is again qualitatively similar to the large, mildly heterogeneous network diagram.

\begin{figure}
    \makebox[0pt][c]{\includegraphics[width=6.5cm, height=6.5cm]{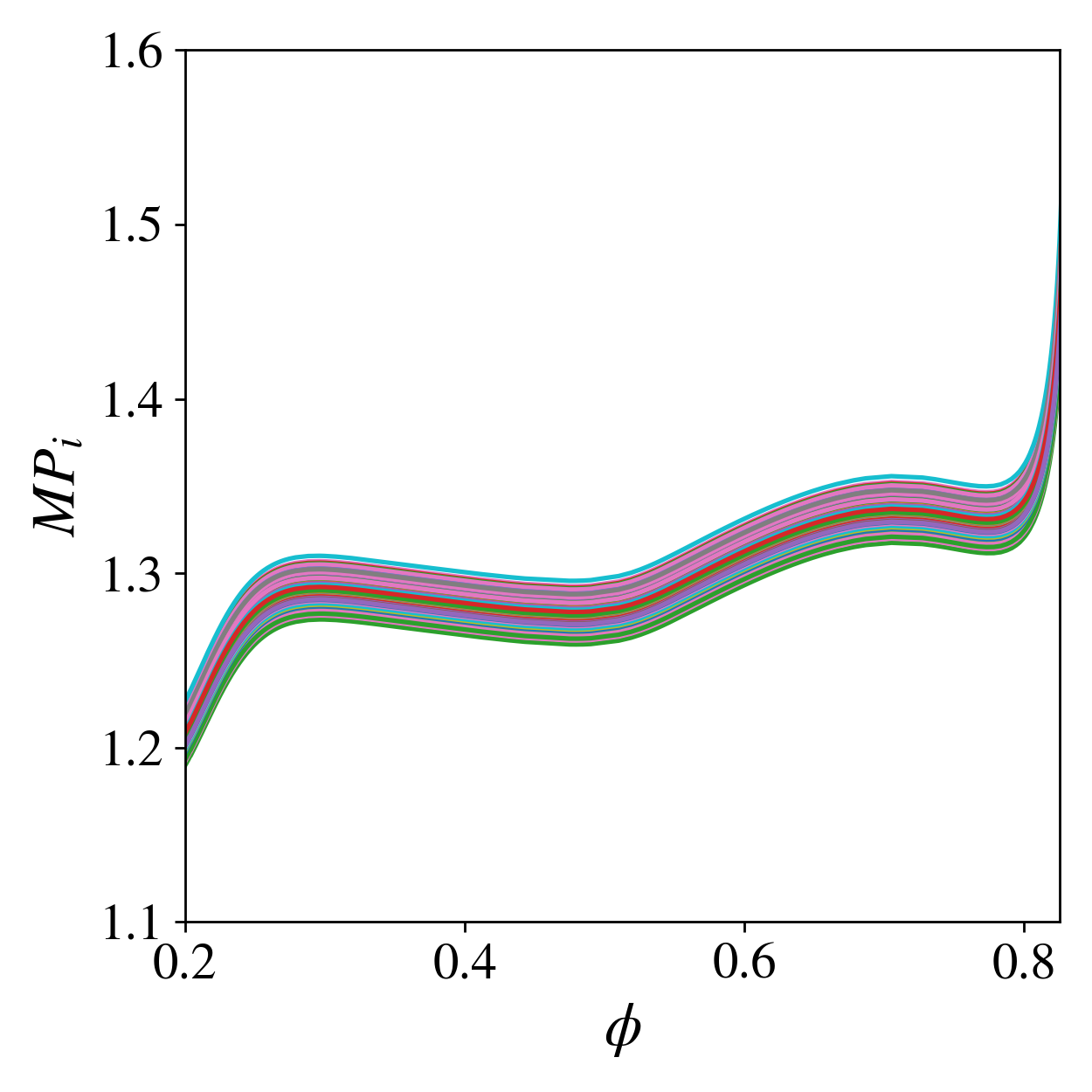}}
    \caption{Continuation  branch of periodic steady states solutions for the neuronal network w.r.t. the duty cycle $\phi$ for  heterogeneity variance of value $\sigma^2=10^{-3}$ and $k_1 =0.49, T_f =24h$. Here the variable $MP_i$ is reported for every neuron $i$ at the initial phase of the forcing. This should be compared to the lower branch in Fig. \ref{fig:phi1}.}
\label{fig:phiN}
\end{figure}

% \begin{figure}
% \begin{subfigure}{.6\textwidth}
%   \centering
%   % include third image
%   \includegraphics[width=1.\linewidth]{Figures/horn2.png}  
% %   \caption{Put your sub-caption here}
%   \label{fig:sub-third}
% \end{subfigure}
% \begin{subfigure}{.4\textwidth}
%   \centering
%   % include fourth image
%   \includegraphics[width=1.\linewidth]{Figures/horn1.png}  
% %   \caption{Put your sub-caption here}
%   \label{fig:sub-fourth}
% \end{subfigure}

% % \caption{Put your caption here}
% \label{fig:fig}
% \end{figure}

\subsection{A Data-Driven Embedding Space for Neuronal Heterogeneity.}

Realistic neuronal networks are characterized by heterogeneity with respect to multiple physical properties. Enumeration and identification of such heterogeneities (and respective parameters for a computational model) from observed neuronal oscillation data is highly nontrivial, and is clearly affected by:  (a) the quality and quantity of data; (b) stochasticity/noise inherent to measurements of biological systems; and (c) the complexity of the computational model and assumptions used in its formulation. 

With increasing model complexity, it is clear that increasing parameter heterogeneity and variability (both for intrinsic kinetic parameters, and for structural network connectivity parameters) becomes important.
When these parameters are not explicitly known, it is the variability of the neuronal behavior itself that encodes it;
and so one can obtain a sense of the parameter heterogeneity from the neuronal time-series variability itself. 
%However, the variability of the oscillation data themselves ought to fully encode the variability 
%of the (heterogeneous) parameter values present in the network. 
In other words, it should be possible to infer the effective parametric heterogeneity/variability of the entire network by the richness of the variability of the observed oscillations of individual neurons. Here, we employ a data-driven approach to uncover an \textit{effective} heterogeneity space purely from observed oscillation data.  
The algorithm we employ is Diffusion Maps (a nonlinear, manifold learning technique), to embed oscillation data in an \textit{effective}, data-driven heterogeneity space \cite{COIFMAN20065, Thiem2021, Kemeth2018}. 
The idea of creating such data-driven {\em emergent
spaces} purely from observations was proposed in  \cite{Kemeth2018,  Kemeth2022}
and it was used in the simple case of Kuramoto-type oscillators in \cite{Thiem2021}.

% \onecolumngrid

% \begin{figure}[H] %12:4.8
%     \centering
%     \includegraphics[width=15cm, height=6cm]{Figures/dmaps2.png}    \caption{(left) Data from a single phase of the limit cycle plotted on the space defined by the first two nontrivial diffusion maps eigenvectors. All datapoints are colored by the \textit{a priori known} heterogeneity parameter $v_{sP_0}$. Data are chosen from the case of $\phi=0.5, k_1 =0.49, \sigma^2 = 10^{-3}$ for forcing angular velocity equal to the intrinsic one. (right) the first nontrivial diffusion maps eigenvector plotted against the heterogeneity parameter.}
%     \label{fig:dmaps}
% \end{figure}
% \twocolumngrid

% \begin{figure*} %12:4.8
%     \centering
%     \includegraphics[width=15cm, height=6cm]{Figures/dmaps2.png}    \caption{(left) Data from a single phase of the limit cycle plotted on the space defined by the first two nontrivial diffusion maps eigenvectors. All datapoints are colored by the \textit{a priori known} heterogeneity parameter $v_{sP_0}$. Data are chosen from the case of $\phi=0.5, k_1 =0.49, \sigma^2 = 10^{-3}$ for forcing angular velocity equal to the intrinsic one. (right) the first nontrivial diffusion maps eigenvector plotted against the heterogeneity parameter.}
%     \label{fig:dmaps}
% \end{figure*}

\begin{figure}[H] %12:4.8
    \centering
    \includegraphics[width=15cm, height=6cm]{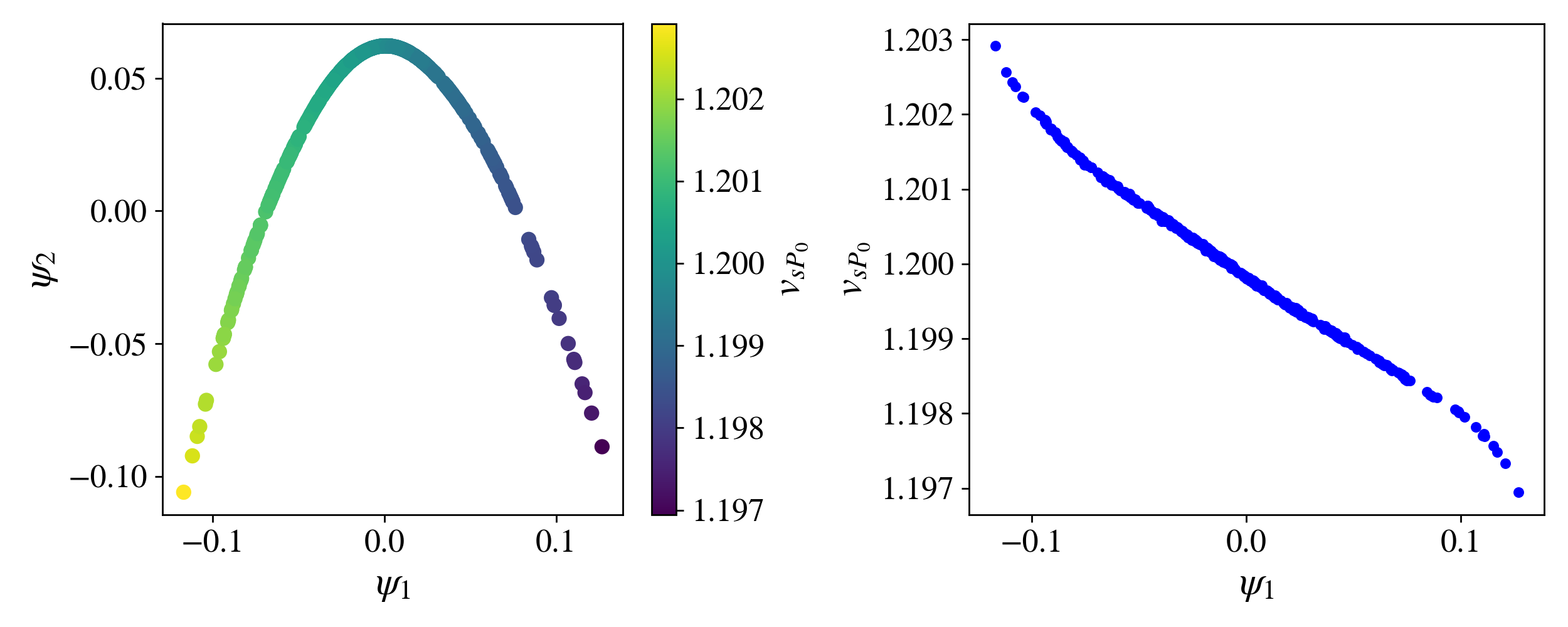}    \caption{(left) Data from a single phase of the limit cycle plotted on the space defined by the first two nontrivial diffusion maps eigenvectors. All datapoints are colored by the \textit{a priori known} heterogeneity parameter $v_{sP_0}$. Data are chosen from the case of $\phi=0.5, k_1 =0.49, \sigma^2 = 10^{-3}$ for forcing angular velocity equal to the intrinsic one. (right) The first nontrivial diffusion maps eigenvector plotted against the heterogeneity parameter.}
    \label{fig:dmaps}
\end{figure}

As in \cite{Kemeth2018} we select data from a single phase of the limit cycle (the exact phase does not matter); the dataset therefore consists of 425 data points (number of neurons) in a 21-dimensional space (number of variables per neuron). As shown in Fig.\ref{fig:dmaps} diffusion maps reveals that the (synchronized) neuronal states are effectively one-dimensional, as $\psi_1$ seems to be the only independent eigenvector. 
The right panel of Fig.\ref{fig:dmaps} confirms that the observed data-based emergent heterogeneity descriptor $\psi_1$ in indeed one-to-one with the (true) intrinsic
heterogeneity $v_{sP_0}$: the data-driven heterogeneity parametrization is one-to-one with the true physical one.

\section{Conclusions}
\label{Sec:conc}

This work was aimed at computationally exploring the limits of entrainment of circadian neurons and circadian neuronal networks. We employ modern techniques from scientific computing, such as matrix-free time-stepper based algorithms, to circumvent limitations inherent in high-dimensional dynamical systems. These algorithms enable the exploration of the simulated effect of (i) forcing frequency, (ii) forcing duty cycle, (iii) Longdaysin dosing level, and (iv) neuronal heterogeneity, on the ability of circadian neurons to entrain to the day/night cycle. Lastly, an unsupervised learning algorithm is used to discover an \textit{effective} neuronal heterogeneity space from observed oscillation data. 

A wealth of responses to different day/night cycle conditions is demonstrated, such as entrainment, quasiperiodicity, phase-locking and chaos. Linking fundamental concepts of nonlinear dynamics to computational neuroscience can lead to a holistic understanding of circadian dynamics and motivate real world applications. Especially in the case of simulated pharmacological effects, such a model can be thought as a computational ``sandbox'' for therapeutic (possibly, personalized) interventions to the SCN. Furthermore, combining scientific computing with machine learning can provide significant insights into the underlying degrees of freedom of systems as complex as the body's own clock.  

\begin{acknowledgments}
This work was partially supported by a US ARO MURI, a NIH U01 grant and the US AFOSR.
\end{acknowledgments}

\onecolumngrid

\twocolumngrid

\nocite{*}
\bibliography{aipsamp}% Produces the bibliography via BibTeX.

\end{document}